\documentclass[a4paper,UKenglish,cleveref, autoref, thm-restate, pdfa, a4paper]{lipics-v2021}

\pdfoutput=1
\hideLIPIcs  

\title{One-Way Communication Complexity of Partial XOR Functions}

\author{Vladimir V. Podolskii}{Tufts University, USA}{vladimir.podolskii@tufts.edu}{https://orcid.org/0000-0001-7154-138X}{}

\author{Dmitrii Sluch}{Nebius, Israel}{dmitrysluch@nebius.com}{https://orcid.org/0009-0004-8363-4101}{}

\authorrunning{V. V. Podolskii and D. Sluch}

\Copyright{Vladimir V. Podolskii and Dmitrii Sluch}



\category{} 

\acknowledgements{We would like to thank the anonymous reviewers for useful comments that helped us improve the presentation.}

\nolinenumbers 

\EventEditors{Karl Bringmann, Martin Grohe, Gabriele Puppis, and Ola Svensson}
\EventNoEds{4}
\EventLongTitle{51st International Colloquium on Automata, Languages, and Programming (ICALP 2024)}
\EventShortTitle{ICALP 2024}
\EventAcronym{ICALP}
\EventYear{2024}
\EventDate{July 8--12, 2024}
\EventLocation{Tallinn, Estonia}
\EventLogo{}
\SeriesVolume{297}
\ArticleNo{79}

\usepackage{tikz}
\usepackage{todonotes}
\usepackage[ruled]{algorithm}
\usepackage{algpseudocode}

\newcommand{\CC}{\mathrm{D_{cc}}}
\newcommand{\CCone}{\mathrm{D_{cc}^{\rightarrow}}}
\newcommand{\DTP}{\mathrm{DT^{\oplus}}}
\newcommand{\NADTP}{\mathrm{NADT^{\oplus}}}
\newcommand{\dist}{\operatorname{dist}}
\newcommand{\Dom}{\operatorname{Dom}}

\newcommand{\poly}{\text{poly}}
\newcommand{\XOR}{\ensuremath{{\rm XOR}}}
\newcommand{\floor}[1]{\lfloor#1\rfloor}

\newcommand{\R}{\mathbb{R}}

\newcommand{\F}{\mathbb{F}}

\definecolor{mc}{rgb}{0.9,0.45,0}
\begin{document}

\maketitle

\begin{abstract}
    Boolean function $F(x,y)$ for $x,y \in \{0,1\}^n$ is an XOR function if $F(x,y) = f(x\oplus y)$ for some function $f$ on $n$ input bits, where $\oplus$ is a bit-wise XOR. XOR functions are relevant in communication complexity, partially for allowing the Fourier analytic technique. For total XOR functions, it is known that deterministic communication complexity of $F$ is closely related to parity decision tree complexity of $f$. Montanaro and Osbourne (2009) observed that one-way communication complexity $\mathrm{D_{cc}^{\rightarrow}}(F)$ of $F$ is exactly equal to non-adaptive parity decision tree complexity $\mathrm{NADT^{\oplus}}(f)$ of $f$. Hatami et al. (2018) showed that unrestricted communication complexity of $F$ is polynomially related to parity decision tree complexity of $f$.

    We initiate the study of a similar connection for partial functions. We show that in the case of one-way communication complexity whether these measures are equal, depends on the number of undefined inputs of $f$. More precisely, if $\mathrm{D_{cc}^{\rightarrow}}(F)=t$ and $f$ is undefined on at most $O\left(\frac{2^{n-t}}{\sqrt{n-t}}\right)$ inputs, then $\mathrm{NADT^{\oplus}}(f)=t$.
    We also provide stronger bounds in extreme cases of small and large complexity.
    
    We show that the restriction on the number of undefined inputs in these results is unavoidable. That is, for a wide range of values of $\mathrm{D_{cc}^{\rightarrow}}(F)$ and $\mathrm{NADT^{\oplus}}(f)$ (from constant to $n-2$) we provide partial functions (with more than $\Omega\left(\frac{2^{n-t}}{\sqrt{n-t}}\right)$ undefined inputs, where $t=\CCone$) for which $\mathrm{D_{cc}^{\rightarrow}}(F) < \mathrm{NADT^{\oplus}}(f)$. In particular, we provide a function with an exponential gap between the two measures. Our separation results translate to the case of two-way communication complexity as well, in particular showing that the result of Hatami et al. (2018) cannot be generalized to partial functions.

    Previous results for total functions heavily rely on the Boolean Fourier analysis and thus, the technique does not translate to partial functions. For the proofs of our results we build a linear algebraic framework instead. Separation results are proved through the reduction to covering codes.
\end{abstract}

\section{Introduction}

In communication complexity model two players, Alice and Bob, are computing some fixed function $F\colon \{0,1\}^n \times \{0,1\}^n \to \{0,1\}$ on a given input $(x,y)$. However, Alice knows only $x$ and Bob knows only $y$. The main object of studies in communication complexity is the amount of communication $\CC(F)$ needed between Alice and Bob to compute the function.

Function $F$ is a XOR-function if for all $x, y \in \{0,1\}^n$ we have $F(x,y) = f(x \oplus y)$ for some $f \colon \{0,1\}^n \to \{0,1\}$, where $x \oplus y$ is a bit-wise XOR of Boolean vectors $x$ and $y$. XOR-functions are important in communication complexity~\cite{ZhangS10,montanaro,TsangWXZ13,Zhang14,ChattopadhyayM17,HatamiHL18,KannanMSY18,AnshuBT19,SinhaW19,Sanyal19,ChattopadhyayMS20,ChattopadhyayGS21,GirishRT22,HambardzumyanHH23,GirishSTW23}, on one hand, since there are important XOR-functions defined based on Hamming distance between $x$ and $y$, and on the other hand, since the structure of XOR-functions allows for the Fourier analytic techniques. In particular, this connection suggests an approach for resolving Log-rank Conjecture for XOR-functions~\cite{ZhangS10,HatamiHL18}.

In recent years there was considerable progress in the characterization of communication complexity of a XOR-function $F$ in terms of the complexity of $f$ in parity decision tree model.
In this model the goal is to compute a fixed function $f$ on an unknown input $x \in \{0,1\}^n$ and in one step we are allowed to query XOR of any subset of input bits. We want to minimize the number of queries that is enough to compute $f$ on any input $x$. The complexity of $f$ in this model is denoted by $\DTP(f)$. It was shown by Hatami et al.~\cite{HatamiHL18} that for any total $f$ we have $\CC(F) = \poly(\DTP(f))$.

Even stronger connection holds for one-way communication complexity case. In this setting only very restricted form of communication is allowed: Alice sends Bob a message based on $x$ and Bob has to compute the output based on this message and $y$. We denote the complexity of $F$ in this model by $\CCone(F)$. The relevant model of decision trees is the model of non-adaptive parity decision trees. In this model we still want to compute some function $f$ on an unknown input and we still can query XORs of any subsets of input bits, but now all queries should be provided at once (in other words, each query cannot depend on the answers to the previous queries). The complexity of $f$ in this model is denoted by $\NADTP(f)$. It follows from the results of  Montanaro, Osbourne~\cite{montanaro} and Gopalan et al.~\cite{gopalan} that for any total XOR-function $F(x,y) = f(x \oplus y)$ we have $\CCone(F) = \NADTP(f)$.

These results on the connection between communication complexity and parity decision trees can be viewed as lifting results. This type of results have seen substantial progress in recent years (see~\cite{RaoY20}).
The usual structure of a lifting result is that we start with a function $f$ that is hard in some weak computational model (for example, a decision tree type model), compose it with some gadget function $g$ to obtain $f \circ g$ (each variable of $f$ is substituted by a copy of $g$ defined on fresh variables) and show that $f \circ g$ is hard in a stronger computational model (for example, a communication complexity type model). The results on XOR-functions can be viewed as lifting results for $g = \XOR$. 

The results on the connection between communication complexity of XOR-functions and parity decision trees discussed above are proved only for total functions $f$ for the reason that the proofs heavily rely on the Fourier techniques. However, in communication complexity and decision tree complexity it is often relevant to consider a more general case of partial functions, and many lifting theorems apply to this type of functions as well, see e.g.~\cite{RezendeMNPRV20,LoffM19,ChattopadhyayMS23,SherstovSW23}. In particular, there are some lifting results for partial functions for gadgets that are stronger than $\XOR$: Mande et al.~\cite{mande} proved such a result for one-way case for inner product gadget (inner product is XOR applied to ANDs of pairs of variables) and Loff, Mukhopadhyay~\cite{LoffM19} proved a result on lifting with equality gadget for general case (note that equality for inputs of length 1 is practically $\XOR$ function). In~\cite{LoffM19} a conjecture is mentioned that for partial XOR-functions $\CC(F)$ is approximately equal to $\DTP(f)$ as well.

\paragraph*{Our results.} In this paper we initiate the studies of the connection between communication complexity for the case of partial XOR functions and parity decision trees. 
It turns out that for one-way case whether they are equal depends on the number of inputs on which the function is undefined: if the number of undefined inputs is small, then the complexity measures are equal and if it is too large, they are not equal. 

More specifically, we show that for $t = \CCone(F)$ the equality $\CCone(F) = \NADTP(f)$ holds if $f$ is undefined on at most $O\left( \frac{2^{n-t}}{\sqrt{n-t}} \right)$ inputs. 
We prove a stronger bound on the number of undefined inputs for small values of $t$. More specifically, for $t=1$ we show that the equality $\CCone(F) = \NADTP(f)$ is true for all partial $f$. For $t=2$ we show that the equality is true for at most $2^{n - 3} - 1$ undefined inputs.
On the other end of the spectrum we show that for any partial function if $\NADTP(f) \geq n-1$, then $\CCone(F) = \NADTP(f)$.

On the other hand, we provide a family of partial functions for which $\CCone(F) < \NADTP(f)$\footnote{Note that the gap in the other direction is impossible: it is easy to see that $\CCone(F) \leq \NADTP(f)$ for all $f$ (see Lemma~\ref{lem:simple-relation} below). Similar inequality (with an extra factor of 2) holds for general communication complexity and parity decision tree complexity.}.
More specifically, we show that for any constant $0 < c< 1$ there is a function $f$ with $\NADTP(f)=cn$ and $\CCone(F) \leq c'n$ for some $c'<c$. 

The number of undefined inputs for the function is $O\left(\frac{2^{dn}}{\sqrt{n}}\right)$ if $c> 1/2$, is equal to $2^{n-1}$ if $c=1/2$, and is $2^n - O\left(\frac{2^{dn}}{\sqrt{n}}\right)$ if $c < 1/2$, where $0<d<1$ is some constant (depending of $c$). 

We provide a function $f$ for which $\NADTP(f) = \sqrt{n \log n}$ and $\CCone(F) \leq O(\log n)$, the number of undefined inputs for $f$ is $2^n - 2^{\Theta(\sqrt{n} \log^{3/2} n)}$. Thus, we provide an exponential gap between the two measures. 

We provide stronger bounds for small and large values of complexity.
For $\CCone(F)=1$ we show that the equality $\CCone(F) = \NADTP(f)$ is true for all partial $f$. For $\CCone(F)=2$ the equality is true for at most $2^{n - 3} - 1$ undefined inputs.
The smallest values of measures for which we provide a separation are $\CCone(F)=7$ and $\NADTP(f)=8$.
On the other end of the spectrum we show that for any partial function if $\NADTP(f) \geq n-1$, then $\CCone(F) = \NADTP(f)$.
The largest value of $\NADTP$ for which we provide a separation is $n-2$, this complements the result that starting with $\NADTP(f)=n-1$ the measures are equal.

All our separation results translate to the setting of two-way communication complexity vs. parity decision trees. In particular, we provide a partial function $f$ with exponential gap between $\CC(F)$ and $\DTP(f)$, which refutes the conjecture mentioned in~\cite{LoffM19}. It is an interesting open problem whether the polynomial relation between these measures discovered by Hatanami et al. for total functions holds for partial functions with some restriction on the number of undefined points.

The techniques behind the results on the connections between communication complexity of XOR-functions and parity decision tree complexity for total functions heavily rely on the Fourier analysis. However, it is not clear how to translate this technique to partial functions. To prove our results, we instead translate the Fourier-based approach of~\cite{montanaro, gopalan} into the language of linear algebra. We design a framework to capture the notion of one-way communication complexity of partial XOR-functions and use this framework to establish equality of $\CCone(F)$ and $\NADTP(f)$ for the small number of undefined points. The separation results can be proved using our framework, but in these version of the paper we provide self-contained proof. The separation results are proved by a reduction to the covering codes.

The rest of the paper is organized as follows. 
In Section~\ref{sec:preliminaries} we provide necessary preliminary information and introduce the notations. 
In Section~\ref{sec:framework} we introduce our linear-algebraic framework.
In Section~\ref{sec:sec_few} we prove main results on the equality of complexity measures. 
In Section~\ref{sec:exponential_gap} we prove separation results.
In Section~\ref{sec:extreme cases} we provide results for extreme cases.
Some of the technical proofs are presented in Appendix.

\section{Preliminaries}
\label{sec:preliminaries}
\subsection{Boolean cube}
A Boolean cube is a graph on the set $\{0, 1\}^n$ of Boolean strings of length $n$. We connect two vertices with an edge if they differ in a single bit only. The set $\{0, 1\}^n$ can also be thought of as the vector space $\F_2^n$, with the bitwise XOR as the group operation. An inner product over this space can be defined as 
\begin{equation}
\langle x, y\rangle = \bigoplus_{i} x_i \wedge y_i.
\end{equation}

Hamming weight of $x$ denoted $|x|$ is defined as the number of coordinates of $x$ equal to $1$. Hamming distance $\dist(x,y)$ between $x \in \{0,1\}^n$ and $y \in \{0,1\}^n$ is the number of coordinates at which $x$ and $y$ differ. The Hamming ball of radius $r$ is a set of vertices of Boolean cube $\{0, 1\}^n$ with Hamming weight not exceeding $r$. We denote by $V(n, r)$ the volume of a Hamming ball in $\{0, 1\}^n$ of radius $r$.

\subsection{Isoperimetric inequalities}
\label{sec:isoperimetric}

\begin{definition}
For a set $A$ we denote the set of neighbors of elements of $A$ as $\Gamma A$. We denote $\Gamma' A := \Gamma A \setminus A$. 
\end{definition}

We will need the vertex isoperimetric inequality for a Boolean cube known as Harper's theorem. To state it we first define Hales order.
\begin{definition}[Hales order~{\cite[Page 56]{harper}}]

\label{def:simplicial_order}
Consider two subsets $x, y \subseteq [m]$ for some natural $m$. We define $x \prec y$ if $|x| < |y|$ or $|x| = |y|$ and the smallest element of the symmetric difference of $x$ and $y$ belongs to $x$. In other words, there exists an $i$ such that $i \in x, i \notin y$, and $i$ is the smallest element in which $x$ and $y$ differ. Here is an example of Hales order for $m = 4$:
\begin{align*}
&\varnothing, \{1\}, \{2\}, \{3\}, \{4\}, \{1, 2\}, \{1, 3\}, \{1, 4\},\{2, 3\}, \{2, 4\}, \{3, 4\}, \\
&\{1, 2, 3\}, \{1, 2, 4\}, \{1, 3, 4\}, \{2, 3, 4\}, \{1, 2, 3, 4\}.
\end{align*}
We can induce Hales order on the set $\{0, 1\}^m$ by identifying subsets of $[m]$ with their characteristic vectors. We define $I_a^m$ to be the set of the first $a$ elements of $\{0, 1\}^m$ in Hales order.
\end{definition}
\begin{theorem}[Harper's theorem~{\cite[Theorem 4.2]{harper}}]
\label{thm:harper}
Let $A \subseteq \{0, 1\}^m$ be a subset of vertices of $m$-dimensional Boolean cube and denote $a = |A|$. Then  $|\Gamma A| \geq |\Gamma I_a^m|$.
\end{theorem}

\subsection{Communication Complexity and Decision Trees}

Throughout this paper, $f$ denotes a partial function $\{0, 1\}^n \to \{0, 1, \perp\}$, we let $\Dom(f) = f^{-1}(\{0,1\})$. We define an XOR-function $F : \{0, 1\}^n \times \{0, 1\}^n \to \{0, 1, \perp\}$ as 
\begin{equation}
F(x, y) = f(x \oplus y).
\end{equation}

In communication complexity model two players, Alice and Bob, are computing some fixed function $F\colon \{0,1\}^n \times \{0,1\}^n \to \{0,1\}$ on a given input $(x,y)$. However, Alice knows only $x$ and Bob knows only $y$. The main subject of studies in communication complexity is the amount of communication $\CC(F)$ needed between Alice and Bob to compute the function. Formal definition of the model can be found in~\cite{nisan}.

We will be mostly interested the in one-way communication model. This is a substantially restricted setting, in which only Alice is permitted to send bits to Bob. Formally, the one-way communication complexity $D_{cc}^{\to}(F)$ is defined to be the smallest integer $t$, allowing for a protocol where Alice knowing her input $x$ sends $t$ bits to Bob, which together with Bob's input $y$ enable Bob to calculate the value of $F$.

The bits communicated by Alice depend only on $x$, that is Alice's message to Bob is $h(x)$ for some fixed total function $h\colon \{0, 1\}^n \to \{0, 1\}^t$. Bob computes the output $F(x, y)$ based on $h(x)$ and his input $y$. That is, Bob outputs $\varphi(h(x),y)$ for some fixed total function $\varphi\colon \{0, 1\}^t \times \{0, 1\}^n \to \{0, 1\}$. If $(x, y)$ is within the domain of $F$, then the equality $\varphi(h(x), y) = F(x, y)$ must be true.

The notion of parity decision tree complexity is a generalization of the well-known decision tree complexity model. In this model, to evaluate a function $f$ for a given input $x$ the protocol queries the parities of some subsets of the bits in $x$. The cost of the protocol on specified input $x$ is the number of queries the protocol makes on that input. The cost of the protocol (sometimes referred to as the worst-case cost) is maximum over all inputs $x$, costs of protocol on the input $x$. The complexity of problem $f$ in the model of parity decision trees $\DTP(f)$ is the minimal over all valid protocols, cost of a protocol for $f$.

We consider the non-adaptive parity decision tree complexity $\NADTP(f)$. This version differs from its adaptive counterpart in that all the queries should be fixed at once. In other words, each next query should not depend on the answers to previous queries. Next, we give a more formal definition of $\NADTP(f)$.

The protocol of complexity $p$ is defined by $n$-bit strings $s_1, \ldots, s_p$ and a total function $l \colon \{0,1\}^p \to \{0,1\}$. On input $x$ the protocol queries the values of
\begin{equation}
\langle s_1, x\rangle, \ldots, \langle s_p, x\rangle
\end{equation}
and outputs 
\begin{equation}
l(\langle s_1, x\rangle, \ldots, \langle s_p, x\rangle).
\end{equation}
The protocol computes partial function $f$, if for any $x \in \Dom(f)$ we have
\begin{equation}
l(\langle s_1, x\rangle, \ldots, \langle s_p, x\rangle) = f(x).
\end{equation}

Throughout the paper $t, h, \varphi, p, s_1, \ldots, s_p, l$ have the same meaning as defined above.

It is easy to see that there is a simple relation between $\NADTP(f)$ and $\CCone(F)$.

\begin{lemma}\label{lem:simple-relation}
    For any $f$ we have $\CCone(F) \leq \NADTP(f)$.
\end{lemma}

\begin{proof}
    Alice and Bob can compute $F(x,y)$ by a simple simulation of $\NADTP$ protocol for $f$. The idea is that they privately calculate the parities of their respective inputs according to $\NADTP$ protocol, then Alice sends the computed values to Bob, who XORs them with his own parities, and then computes the value of $F$. 

    More formally, assume that $\NADTP(f)=p$ and the corresponding protocol is given by $s_1, \ldots, s_p \in \{0,1\}^n$ and a function $l$, that is
    \begin{equation}
    \forall x \in \Dom(f), f(x) = l(\langle s_1, x \rangle, \ldots, \langle s_p, x\rangle).
    \end{equation} 
    
    For $i \in [p]$, we let
    \begin{equation}
    h_i(x) := \langle s_i, x \rangle.
    \end{equation}
    For the communication protocol of complexity $p$ we let
    \begin{align}
    &h(x) = (h_1(x), \ldots, h_p(x)),\\
    &\varphi(a, y) := l(a_1 \oplus \langle s_1, y \rangle, \ldots, a_p \oplus \langle s_p, y\rangle).
    \end{align}

    Then for any $(x, y)$ such that $x \oplus y \in \Dom(f)$ we have 
\begin{align}
        \varphi(h(x), y) =\ & l(h_1(x) \oplus \langle s_1, y\rangle, \ldots, h_p(x) \oplus \langle s_p, y\rangle) = \\
        & l(\langle s_1, x\rangle \oplus \langle s_1, y\rangle, \ldots, \langle s_p, x\rangle \oplus \langle s_p, y\rangle) = \\
        & l(\langle s_1, x \oplus y \rangle, \ldots, \langle s_p, x \oplus y \rangle) = f(x \oplus y) = F(x, y).
\end{align}

We constructed a $p$-bit communication protocol for $F$, and thus 
\begin{equation}
\CCone(F) \leq p = \NADTP(f).
\end{equation}
\end{proof}

In this paper, we are mainly interested in whether the inequality in the opposite direction is true.

\subsection{Covering Codes}
\label{sec:covering-codes}

\begin{definition}
    A subset $\mathcal{C} \subseteq \{0,1\}^n$ is a $(n,K,R)$ covering code if $|\mathcal{C}| \leq K$ and for any $x \in \{0,1\}^n$ there is $c \in \mathcal{C}$ such that $\dist(x,c) \leq R$. In other words, all points in $\{0,1\}^n$ are covered by balls of radius $R$ with centers in $\mathcal{C}$.
\end{definition}

The following general bounds on $K$ are known for covering codes.

\begin{theorem}[{\cite[Theorem 12.1.2]{CohenHLL97}}]
    \label{thm:covering-codes}
    For any $(n,K,R)$ covering code we have
    \begin{equation}
    \log K \geq n - \log V(n,R).
    \end{equation}
    For any $n$ and any $R \leq n$ there is a $(n,K,R)$ covering code with
    \begin{equation}
    \log K \leq n - \log V(n,R) + \log n.
    \end{equation}
\end{theorem}

We will use the following well known fact.
\begin{theorem}[{\cite[Section 2.6]{CohenHLL97}}]
    \label{thm:hamming-codes}
    If $n = 2^m - 1$ for some $m$, then Boolean cube $\{0,1\}^n$ can be splitted into disjoint balls of radius 1. 
\end{theorem}

This construction is known as a Hamming error correcting code. Note that it is a $(n = 2^m -1,\frac{2^n}{n+1},1)$ covering code.

\begin{definition}
    For two covering codes $\mathcal{C}_1$ and $\mathcal{C}_2$ their direct sum is 
    \begin{equation}
    \mathcal{C}_1 \oplus \mathcal{C}_2 = \{(c_1, c_2) \mid c_1 \in \mathcal{C}_1, c_2 \in \mathcal{C}_2\}.
    \end{equation}
\end{definition}

\begin{lemma}[{\cite[Theorem 12.1.2]{CohenHLL97}}] \label{lem:direct-sum}
    If $\mathcal{C}_1$ is a $(n_1, K_1, R_1)$ covering code and $\mathcal{C}_2$ is a $(n_2, K_2, R_2)$ covering code, then $\mathcal{C}_1 \oplus \mathcal{C}_2$ has parameters $(n_1 + n_2, K_1K_2, R_1 + R_2)$.
\end{lemma}

We need the following bounds on the sizes of Hamming balls (see, e.g.~\cite[Appendix A]{Jukna12}).

\begin{lemma} \label{lem:binomial-universal}
    For any $n$ and  $k \leq n$ we have
    \begin{equation}
    \left( \frac{n}{k}\right)^k \leq  V(n,k) \leq \left( \frac{en}{k}\right)^k.
    \end{equation}
\end{lemma}

\begin{lemma} \label{lem:binomial-linear-case}
    For any constant $0 < c < 1$ we have
    \begin{equation}
    \binom{n}{cn} = O\left( \frac{1}{\sqrt{n}}2^{H(c)n}\right).
    \end{equation}
    For any constant $0 < c < 1/2$ we have
    \begin{equation}
    V(n,cn) = O\left( \frac{1}{\sqrt{n}}2^{H(c)n}\right),
    \end{equation}
    where $H$ is the binary entropy function.
\end{lemma}

\begin{lemma}[{\cite[Section 5.4]{Spencer-book14}}] \label{lem:spencer}
    \begin{equation}
    V\left(n,\frac{n}{2} - \Theta(\sqrt{n \log n})\right) =  \frac{2^{n}}{\poly(n)}.
    \end{equation}
\end{lemma}

For the binary entropy function $H(x)$ we will use the following simple fact.

\begin{lemma} \label{lem:entropy-convergence}
    For any constant $c \in (0,1)$ and for any $\alpha_n \xrightarrow[n \to \infty]{} 0$ we have
    \begin{equation}
    H(c + \alpha_n) = H(c) + O(\alpha_n),
    \end{equation}
    where the constant in $O$-notation might depend on $c$, but not on $n$.
\end{lemma}

This is true since the derivative of $H$ is upper bounded by a constant in any small enough neighborhood of $c$.

\section{Linear-algebraic framework}
\label{sec:framework}

\subsection{Graph-based analysis of one-way communication protocols}
Recall that in a one-way communication protocol of complexity $t$ for $F(x,y) = f(x\oplus y)$ Alice on input $x \in \{0,1\}^n$ first sends to Bob $h(x)$ for some fixed $h \colon \{0,1\}^n \to \{0,1\}^t$. After that Bob computes the output $\varphi(h(x),y)$, where $y \in \{0,1\}^n$ is Bob's input and $\varphi \colon \{0,1\}^t \times \{0,1\}^n \to \{0,1\}$.

Let's consider the partition $\mathcal{H} = \{H_a \mid a \in \{0,1\}^t\}$, where for any $a \in \{0, 1\}^{t}$
\begin{equation}
H_{a} = h^{-1}(a).
\end{equation}
We refer to $\mathcal{H}$ as \emph{$h$-induced partition}. A class $H_a$ of this partition is the set of inputs for which Alice sends Bob the same message. 

Consider two arbitrary inputs $x, y \in \{0, 1\}^n$.
We call the vector $\Delta = x \oplus y$ the \emph{shift} between $x$ and $y$. The intuition is that $y$ is equal to the shift $x \oplus \Delta$ of $x$ by $y$ (and vise versa). 

We say that $\Delta \in \{0,1\}^n$ is a \emph{good shift} if there is a pair $x,y \in \{0,1\}^n$ such that $x \oplus y = \Delta$ and $h(x) = h(y)$, or equivalently, if $x$ and $y$ belong to the same class of $\mathcal{H}$. 
Note that $f$ does not necessarily need to be defined on inputs $x$ and $y$. 
However, it turns out that on the domain of $f$ the value of $f$ is invariant under good shifts.

\begin{lemma} \label{lem:good-shift-main-property}
    Assume that $\Delta$ is a good shift. Consider any $v, u \in \Dom(f)$ such that $v \oplus u = \Delta$. Then, $f(v)=f(u)$.
\end{lemma}

\begin{proof}
    Since $\Delta$ is good, there are $x$ and $y$ such that $h(x) = h(y)$ and $x\oplus y = \Delta$. Then
    \begin{equation}
    f(v) = \varphi(h(x), x \oplus v) = \varphi(h(y), x \oplus v) = f(v \oplus x \oplus y) = f(v \oplus \Delta) = f(u).
    \end{equation}
\end{proof}

\begin{figure}[t]
\centering
	\begin{tikzpicture}
		\path[use as bounding box,help lines,draw=none] (-1,-1) grid (8.2,3);
		\foreach \x/\y in {0/0, 0/2, 2/0, 2/2}{
			\draw[thick] (\x,\y,0) -- (\x,\y,2);
			}
		\foreach \x/\z in {0/0, 0/2, 2/0, 2/2}{
			\draw[thick] (\x,0,\z) -- (\x,2,\z);
		}
		\foreach \y/\z in {0/0, 0/2, 2/0, 2/2}{
			\draw[thick] (0,\y,\z) -- (2,\y,\z);
		}
		\node (h) at (1,2.5) {$h \colon \{0, 1\}^3 \to \{a,b,c,d,e\}$};
		\foreach \x/\y/\z/\l/\v in {0/0/0/left/{\color{mc}c}, 0/0/2/left/{\color{mc}a}, 0/2/0/left/d, 
		2/0/0/right/{\color{mc}c}, 2/2/0/right/{\color{mc}b}, 2/0/2/right/e, 
		0/2/2/left/{\color{mc}a}, 2/2/2/right/{\color{mc}b}, } {
			\node[\l] at (\x,\y,\z) {$\v$};
		}
		\draw[very thick,mc] (0,0,2) -- (0,2,2);
		\draw[very thick,mc] (2,2,0) -- (2,2,2);
		\draw[very thick,mc] (0,0,0) -- (2,0,0);
		\draw[thick] (2,0,2) -- (2,2,2);
		\node (Ghf) at (7,2.5) {$G_{h}$};
		\foreach \x/\y in {6/0, 6/2, 8/0, 8/2}{
			\draw[thick] (\x,\y,0) -- (\x,\y,2);
		}
		\foreach \x/\z in {6/0, 6/2, 8/0, 8/2}{
			\draw[thick] (\x,0,\z) -- (\x,2,\z);
		}
		\foreach \y/\z in {0/0, 0/2, 2/0, 2/2}{
			\draw[thick] (6,\y,\z) -- (8,\y,\z);
		}
	\end{tikzpicture}
 \caption{Example of total h-induced graph}
 \label{fig:total_h_induced_graph}
\end{figure}
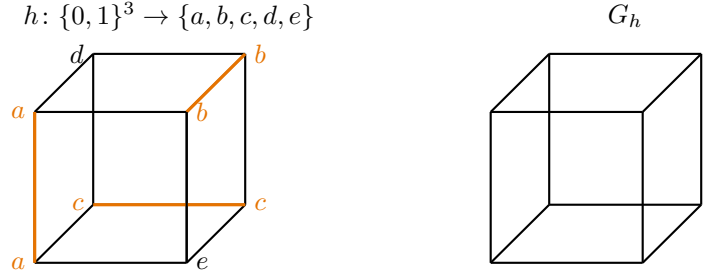

This leads us to the following notion.
\begin{definition}

For the functions $f: \{0, 1\}^n \to \{0, 1\}$, $f(x \oplus y) = \varphi(h(x), y)$ 
let the \emph{total $h$-induced graph} be the graph with vertices $\{0, 1\}^n$ and with an edge between $x \in \{0, 1\}^n$ and $y \in \{0, 1\}^n$ if $x \oplus y$ is a good shift for $h$. Now remove vertices where the function $f$ is undefined. The resulting graph is called the \emph{partial $h$-induced graph}.

There is an alternative way of thinking about total $h$-induced graph. Consider a graph with vertices labeled $\{0, 1\}^n$ in which we connect two vertices if the value of $h$ on these vertices is the same. Clearly it is a subgraph of the total $h$-induced graph. Now consider a shift of this graph, that is, a graph in which we XORed labels of all vertices with some fixed vector. This graph is also a subset of the total $h$-induced graph. By considering all possible shifts and taking the union of all graphs we will get the total $h$-induced graph. See Figure~\ref{fig:total_h_induced_graph} for an example of total h-induced graph.
\end{definition}

By transitivity, if $(h, \varphi)$ form a valid communication protocol then $f$ assigns identical values to each connected component in partial $h$-induced graph. The converse is also true.
\begin{theorem}

\label{thm:ccone_to_graph_reduction}
Consider a function $f: \{0, 1\}^n \to \{0, 1\}$. For a function $h: \{0, 1\}^n \to \{0, 1\}^t$ there is a function $\varphi: \{0, 1\}^t \times \{0, 1\}^n \to \{0, 1\}$ such that $(h, \varphi)$ form a valid communication protocol for $f$ if and only if $f$ assigns the same value to each connected component in the partial $h$-induced graph.
\end{theorem}
\begin{proof}
As discussed above, if $(h, \varphi)$ forms a valid communication protocol, then $f$ assigns the same value to each connected component of the partial $h$-induced graph. 

It remains to prove the converse statement.
We assume that $f$ assigns the same value to each connected component and we need to show that there is $\varphi$ such that 
\begin{equation}
\forall (x, y) \in \Dom(F),\ \  F(x, y) = \varphi(h(x), y).
\end{equation} 
The proof idea is the following. Each input $(x, y) \in \{0, 1\}^n \times \{0, 1\}^n$ to $F$ yields an input $(a, y) \in \{0, 1\}^t \times \{0, 1\}^n$ to $\varphi$ where $\alpha = h(x)$. We define $\varphi$ on $(\alpha, y)$ to be equal to $F$ on a single corresponding $F$-input $(x', y)$ with $x' \in h^{-1}(\alpha)$. Then we prove that $\varphi$ defined that way gives a communication protocol computing $F$ correctly on all inputs $(x, y) \in \Dom(f)$

Formally, we define $\varphi$ as follows.
For each $\alpha \in \{0,1\}^t$ and $y \in \{0,1\}^n$, consider $x' \in \{0,1\}^n$ such that $h(x') = \alpha$ and $(x', y) \in \Dom(F)$. 
If there is no such $x'$ we define $\varphi(\alpha, y)$ arbitrarily.
If there is such an $x'$, let 
\begin{equation}
\varphi(\alpha, y) := F(x', y).
\end{equation}

Now we show that the resulting protocol computes $F(x,y)$ correctly for any $(x,y)$. 

Consider arbitrary $(x, y) \in \Dom(F)$. 
Consider $x'$ chosen for $\alpha = h(x)$ and $y$ (it exists, since clearly $x$ itself satisfies all the necessary conditions).

Thus, we have
\begin{equation}
\varphi(h(x), y) = F(x', y).
\end{equation}
It remains to prove that 
\begin{equation}
F(x', y) = F(x, y)
\end{equation}
or equivalently,
\begin{equation}
f(x' \oplus y) = f(x \oplus y).
\end{equation} 
For XOR of these two inputs of $f$ we have 
\begin{equation}
(x' \oplus y) \oplus (x \oplus y) = x' \oplus x.
\end{equation}
Since $h(x) = h(x')$, we have that $x' \oplus x$ is a good shift. And since 
\begin{equation}
(x, y), (x', y) \in \Dom(F),
\end{equation}
we have 
\begin{equation}
x \oplus y, x' \oplus y \in \Dom(f).
\end{equation} 
 
We have that vertices $x \oplus y$ and $x'\oplus y$ are connected in the partial $h$-induced graph and by Lemma~\ref{lem:good-shift-main-property} $f$ assigns the same value to them. Hence, the function $\varphi$, together with $h$, forms a communication protocol for $F$.
\end{proof}
\subsection{Using coset structures on a Boolean cube to analyze non-adaptive parity decision trees.}

We consider the vertices of the Boolean cube as a vector space $\F_2^n$. We show that a $\NADTP$ protocol corresponds to a linear subspace of $\F_2^n$ such that $f$ is constant on each of its cosets (the coset for a linear subspace $L$ and a vector $l$ is defined as the set $\{x + l | l \in L\}$ and denoted $L + l$).
\begin{theorem}
\label{thm:napdt_to_cosets_reduction}
Let $f \colon \{0,1\}^n \to \{0,1\}$.
There is a $p$-bit $\NADTP$ protocol for $f$ if and only if there exists an $n - p$ dimensional subspace of $\{0, 1\}^n$ such that for each coset of that subspace, $f$ assigns the same value to all inputs of the coset where $f$ is defined.
\end{theorem}
\begin{proof}
Suppose $s_1, \ldots, s_p, l$ form a valid $\NADTP$ protocol for $f$. We construct a matrix $S$ with rows $s_1, \ldots, s_p$. If some of the rows are linearly dependent, we add rows arbitrarily to make the rank of $S$ equal to $p$. When $S$ is multiplied on the right by some vector $x$, we obtain all inner products of $x$ with vectors $s_1, \ldots, s_p$ (and possibly other bits if we added rows). 

Consider the vector subspace $\{x | Sx = 0\}$. This is an $n - p$ dimensional space. For all points in the same coset of this subspace, the tuple consisting of values of the inner products $(\langle s_1, x \rangle, \ldots, \langle s_p, x \rangle)$ is the same, so is the value of $l(\langle s_1, x \rangle, \ldots, \langle s_p, x \rangle)$. For all points where $f$ is defined and lying in the same coset, the value of $f$ must be equal to the value of $l$ and thus the same for all points in the coset.

In the reverse direction, let $\langle e_1, \ldots, e_{n - p} \rangle$ be an $n - p$ dimensional subspace of $\{0, 1\}^n$ such that for each of its cosets $f$ is constant on all points of that coset on which it is defined. We can represent this subspace in the form $\{x | Sx = 0\}$ for some matrix $S$ of size $p \times n$.

Vectors $x$ and $y$ are in the same coset of $\langle e_1, \ldots, e_{n - p} \rangle$ iff $Sx = Sy$.
Thus, to compute $f(x)$ it is enough to compute 
the inner product of $x$ with the rows of $S$.

\end{proof}
\begin{corollary}
\label{cor:bounding_nadt_through_ccone}

Consider a function $f: \{0, 1\}^n \to \{0, 1\}$ having valid communication protocol $f(x \oplus y) = \varphi(h(x), y)$ where $h : \{0, 1\}^n \to \{0, 1\}^t, \varphi: \{0, 1\}^t \times \{0, 1\}^n \to \{0, 1\}$. Suppose there is an $n - t$ dimensional subspace $L$ of $\{0, 1\}^n$ and consider subgraphs of partial $h$-induced graph each over vertices belonging to different cosets of $L$. If all of these subgraphs are connected then $\NADTP(f) \leq t$.
\end{corollary}
\begin{proof}
By Theorem~\ref{thm:ccone_to_graph_reduction} $f$ is constant on each coset. By Theorem~\ref{thm:napdt_to_cosets_reduction} it follows that $\NADTP(f) \leq t$.
\end{proof}
\section{Equality between \texorpdfstring{$\CCone(F)$}{one-way communication complexity} and \texorpdfstring{$\NADTP(f)$}{parity decision tree complexity}}
\label{sec:sec_few}

In this section we will show that if $\CCone(F)=t$ and the number of undefined inputs is small, then $\NADTP(f)=t$ as well. More specifically, we prove the following theorem.
\begin{theorem}
    \label{theorem:few_undefined_points}
    If for the function $f \colon \{0,1\}^n \to \{0,1\}$ we have $\CCone(F)=t$, where $F(x, y) = f(x \oplus y)$, and $f$ is undefined on less than $\binom{n - t + 1}{\lfloor{\frac{n - t}{2}}\rfloor - 1}$ inputs, then $\NADTP(f) = t$.
\end{theorem}

By Lemma~\ref{lem:binomial-linear-case} we have that $\binom{n-t+1}{\left\lfloor\frac{n-t+1}{2}\right\rfloor} = O(\frac{2^{n - t}}{\sqrt{n - t}})$ and since $\lfloor{\frac{n - t}{2}}\rfloor - 1$ differs from $\left\lfloor\frac{n-t+1}{2}\right\rfloor$ by only a constant, it is easy to see that the same estimate applies to $\binom{n - t + 1}{\lfloor{\frac{n - t}{2}}\rfloor - 1}$ as well. Thus, the number of undefined inputs is $O(\frac{2^{n - t}}{\sqrt{n - t}})$.

The rest of the section is devoted to the proof of Theorem~\ref{theorem:few_undefined_points}.
The idea of the proof is as follows.
Consider the $h$-induced partition $\mathcal{H}$ corresponding to the communication protocol of complexity $t$. We show that either the partition $\mathcal{H}$ corresponds to the cosets of an $n - t$ dimensional subspace of $\{0, 1\}^n$, which allows us to construct an $\NADTP$ protocol, or there exist \emph{many} good shifts. The structure of these good shifts imposes restrictions on $f$ that again allow us to construct an $\NADTP$ protocol.

We start with a simple case.
\begin{lemma} 
\label{lemma:h_cosets_imply_small_napdt} 
If there exists $t$-bit communication protocol, $(h, \varphi)$ for a function $f \colon \{0,1\}^n \to \{0,1\}$, and the $h$-induced partition $\mathcal{H}$ corresponds to cosets of an $n - t$ dimensional subspace $L$ of $\{0, 1\}^n$, then $\NADTP(f) \leq t$.
\end{lemma}
\begin{proof}
Since the partition $\mathcal{H}$ corresponds to the cosets of $L$, we have that for any inputs $x$ and $y$, if $h(x)=h(y)$, then $x \oplus y \in L$ and vice versa. In other words, all good shifts are in $L$ and any shift in $L$ is good. Thus, connected components of the total $h$-induced graph are cosets of $L$ and are fully connected. By Corollary~\ref{cor:bounding_nadt_through_ccone} we have that $\NADTP(f) \leq t$.
\end{proof}

The structure of the proof for the other case is the following. We show that the total $h$-induced graph is structured into connected components, each of which is a coset of a $k$-dimensional subspace of $\{0, 1\}^n$ for $k \geq n - t$. We show that there is a bijective graph homomorphism of the $k$-dimensional Boolean cube onto each component. Furthermore, each vertex in the total $h$-induced graph has a degree of at least $\frac{2^n}{2^t} - 1$. We show that if we remove fewer than $\binom{n - t + 1}{\floor{\frac{n - t}{2}} - 1}$ vertices, each coset still contains one connected component. By the way of contradiction, suppose this is not the case and some coset contains more than one connected component. We consider the smallest of these components, denote the set of its nodes by $S$. We show that the number of neighboring vertices of $S$ in the total $h$-induced graph (excluding $S$ itself) is not less than $\binom{n - t + 1}{\floor{\frac{n - t}{2}} - 1}$. This implies that after removing the undefined inputs of $f$ $S$ cannot not be separated from other nodes in the coset. To show this we treat separately cases of large and small $|S|$. For small $|S|$ we use the fact that vertices have high degree. For large $|S|$ we use the vertex-isoperimetric inequality for the Boolean cube.

\begin{lemma}
\label{lemma:large_shifts_dimension}
    Suppose there exists a $t$-bit communication protocol $(h, \varphi)$ for $f\colon \{0,1\}^n \to \{0,1\}$ and the $h$-induced partition $\mathcal{H}$ classes do not correspond to cosets of an $n - t$-dimensional subspace of $\{0, 1\}^n$. Let $D$ be the set of good shifts for $h$. Then $D$ contains a minimum of $n - t + 1$ linearly independent vectors.
\end{lemma}
\begin{proof}
    Suppose there are at most $n - t$ linearly independent good shifts $e_1, \ldots, e_{n - t}$. Consider a linear subspace of $\{0, 1\}^n$ spanned over by these shifts and add some vectors to it to make it exactly $n - t$ dimensional if needed. Denote the resulting subspace $L$. As classes of $\mathcal{H}$ do not correspond to the cosets of $L$ and there are $2^{t}$ of both classes and cosets there exist two elements belonging to the same class and different cosets. Their XOR is a good shift linearly independent with $e_1, \ldots, e_{n - t}$. We got a contradiction implying the lemma.
\end{proof}

\begin{lemma}
    \label{lemma:cube_structure}
    Suppose there exists $t$-bit communication protocol $(h, \varphi)$ for $f$. Let $D$ be the set of all good shifts for $h$ and $\{ e_1, \ldots, e_k \}$ be the largest linearly independent subset of $D$. Then the total $h$-induced graph $\mathcal{H}$ has the following properties.
    \begin{itemize}
        \item Cosets of the subspace $\langle e_1, \ldots, e_k\rangle$ are connected components of $\mathcal{H}$.
        \item There is a bijective graph homomorphism of $k$-dimensional Boolean cube into each coset.
    \end{itemize} 
\end{lemma}
\begin{proof}
    It is easy to see that all vertices in any coset are connected to each other.
    Let's show that no edges exist between vertices of different cosets. Assume by contradiction that there is an edge between vertices $v$ and $u$ from different cosets. Note that $u \oplus v \notin \langle e_1, \ldots, e_k \rangle$. 
    Thus, vectors $e_1, \ldots, e_k, u \oplus v$ form a linearly independent system of size $k+1$, which is a contradiction.
    
    Now, let's construct a homomorphism $q$ from the Boolean cube $\{0,1\}^k$ into the coset $v + \langle e_1, \ldots, e_k \rangle$ for an arbitrary vertex $v$. Consider a matrix $B$ that has vectors $e_1, \ldots, e_k$ as its columns and let $q(x) = v \oplus Bx$. The image of $q$ is within the coset $v + \langle e_1, \ldots, e_k \rangle$, as columns of $B$ belong to the subspace $\langle e_1, \ldots, e_k \rangle$. The mapping is bijective on $v + \langle e_1, \ldots, e_k \rangle$, as $B$'s columns are linearly independent. Finally, consider a pair of vertices $x, y$ adjacent in a Boolean cube. Since the vertices are adjacent, they only differ in a single bit $i$. Thus, 
    \begin{equation}
    q(x) \oplus q(y) = (v \oplus Bx) \oplus (v \oplus By) = B(x \oplus y) = e_i.
    \end{equation}
    Since $e_i \in D$, an edge exists between $q(x)$ and $q(y)$, implying that $q$ is a graph homomorphism.
\end{proof}

\begin{lemma}
    \label{lemma:vertex_degree}
    Suppose there exists $t$-bit communication protocol $(h, \varphi)$ for $f \colon \{0,1\}^n \to \{0,1\}$. Then in the total $h$-induced graph, the degree of any vertex is not less than $\frac{2^n}{2^t} - 1$.
\end{lemma}
\begin{proof}
    Let's consider the largest class in the $h$-induced partition $\mathcal{H}$. Since the number of classes is at most $2^t$, the largest class contains at least $\frac{2^n}{2^t}$ elements. Fix an element of the class and compute its XOR with all elements in the same class $\mathcal{H}$. We have $\frac{2^n}{2^t}$ XORs in total, $\frac{2^n}{2^t} - 1$ of which are non-zero. Since each XOR is computed between elements in the same class, these XORs are good shifts. For all vertices in the $h$-induced graph for each good shift we draw an edge from the vertex corresponding to this shift. Therefore, the degree of any vertex is at least $\frac{2^n}{2^t} - 1$.
\end{proof}

\begin{lemma}
\label{lemma:isoperim_final}
  If $A$ is a subset of $k$-dimensional Boolean cube satisfying $V \left(m, \left\lfloor\frac{m - 1}{2}\right\rfloor - 2\right) \leq |A| \leq 2^{k - 1}$ for some $m$, then $|\Gamma' A| \geq \binom{m}{\left\lfloor\frac{m - 1}{2}\right\rfloor - 1}$.
\end{lemma}
The proof of the lemma is moved to Appendix~\ref{app:isoperim}. Finally, we are ready to prove Theorem~\ref{theorem:few_undefined_points}.
\begin{proof}[Proof of Theorem~\ref{theorem:few_undefined_points}]
    We are given $t$-bit communication protocol $(h, \varphi)$ for $F$. By Lemma~\ref{lemma:large_shifts_dimension}, the $h$-induced partition $\mathcal{H}$ either corresponds to cosets of an $n - t$ dimensional subspace of $\{0, 1\}^n$ (and then by Lemma~\ref{lemma:h_cosets_imply_small_napdt} we have $\NADTP(f) \leq t$), or the set of good shifts $D$ contains at least $n - t + 1$ linearly independent vectors. Let $\langle e_1, \ldots, e_k\rangle$, where $k \geq n - t + 1$, be the largest subset of linearly independent vectors in $D$. Consider the cosets of the subspace $\langle e_1, \ldots, e_k\rangle$.
    We will show that if we remove fewer than $\binom{n - t + 1}{\lfloor{\frac{n - t}{2}}\rfloor - 1}$ vertices from the total $h$-induced graph, each coset will contain no more than one connected component.
    Assume by contradiction that after removing the vertices, some coset splits into several connected components. Let $A$ be the smallest of these components. If there are at most $V(n - t + 1, \lfloor{\frac{n - t}{2}}\rfloor - 2) - 1$ vertices in $A$, consider a vertex $a$ in $A$. Given the degree of $a$ is at least $2^{n - t} - 1$, $a$ has at least 
    \begin{multline}
    2^{n - t} - V\left(n - t + 1, \left\lfloor{\frac{n - t}{2}}\right\rfloor - 2\right)\\ \geq V\left(n - t + 1, \left\lfloor\frac{n - t}{2}\right\rfloor\right) - V\left(n - t + 1, \left\lfloor{\frac{n - t}{2}}\right\rfloor - 2\right) \geq \binom{n - t + 1}{\left\lfloor{\frac{n - t}{2}}\right\rfloor - 1} 
    \end{multline}
    neighbors outside $A$.
    
    On the other hand, suppose $A$ has at least $V(n - t + 1, \lfloor{\frac{n - t}{2}}\rfloor - 2)$ vertices. Since $A$ is the smallest connected component in its coset it also follows that $A$ has no more than $2^{k - 1}$ vertices. 
    By Lemma~\ref{lemma:isoperim_final} we have $|\Gamma' A| \geq \binom{n - t + 1}{\left\lfloor\frac{n - t}{2}\right\rfloor - 1}$, which is more than the number of removed vertices, a contradiction. Thus, cosets cannot be split into several components and by Corollary~\ref{cor:bounding_nadt_through_ccone} we have $\NADTP(f) \leq n - k \leq t - 1$, which is a contradiction.
\end{proof}

\section{Separations between \texorpdfstring{$\CCone(F)$}{owCC} and \texorpdfstring{$\NADTP(f)$}{PDT}}
\label{sec:exponential_gap}

In this section we show that if the number of undefined inputs is large, there is a gap between $\CCone(F)$ and $\NADTP(f)$. That is, we aim to come up with a function $f$ such that $\CCone(F)$ is small and $\NADTP(f)$ is large.

The key idea in our construction is that in $h$-induced graph for the intended communication protocol the edges connect only vertices with small Hamming distance between them. Then, if the function $f$ has 0-inputs and 1-inputs far away from each other, they are not connected and $h$ corresponds to a valid protocol. We will ensure that at the same time $f$ has large $\NADTP$ complexity.

We start with the construction of the functions, then investigate their $\NADTP$ complexity and then prove upper bounds on $\CCone$ complexity of the corresponding XOR functions. The latter part is through the reduction to covering codes.

\begin{definition}
    For a parameter $k$ define $f_k \colon \{0,1\}^n \to \{0,1,\perp\}$ in the following way.
    \begin{equation}
    f_k(x) = \begin{cases}
        0 & \text{for } |x| \leq k,\\
        \perp & \text{for } k+1 \leq |x| \leq n-1, \\
        1 & \text{for } |x| = n.
    \end{cases}
    \end{equation}
    We denote the corresponding XOR function by $F_k$.
\end{definition}

Note, that the number of undefined inputs in $f_k$ is $V(n,n-k-1) - 1$. 

It turns out that $f_k$ has reasonably large $\NADTP$ and $\DTP$ complexities.

\begin{theorem}
\label{thm:counterexample_dtp_large}
$\NADTP(f_k) = \DTP(f_k) = k + 1$.
\end{theorem}
\begin{proof}

Since $\DTP(f) \leq \NADTP(f)$ for any $f$, it is enough to prove that $\NADTP(f_k) \leq k+1$ and $\DTP(f_k) \geq k+1$.

For the upper bound, observe that it is enough to query variables $x_1, \ldots, x_{k+1}$. If all of them are equal to 1, we output 1, otherwise we output 0. It is easy to see that this protocol computes $f_k$ correctly.

For the lower bound suppose, for the sake of contradiction, that an adaptive parity decision tree exists that can compute the function $f$ with $k$ or fewer queries. Consider the path corresponding to the input $e = (1,\ldots,1)$. Let's assume that the decision tree queried the parities $\langle s_i, e\rangle$ for $s_1, \ldots, s_k$. The answers to the queries are equal to $\langle s_1, e \rangle, \ldots, \langle s_k, e\rangle$. Consider a matrix $B \subseteq \mathbb{F}^{k\times n}$ consisting of rows $s_1, \ldots, s_k$. 

Denote $a = B e$. In particular, we have that $a$ lies in the subspace generated by columns of $B$. Since the rank of $B$ is at most $k$ (the matrix has $k$ rows), there is a subset of at most $k$ columns generating this subspace. In particular, there is $x \in \{0,1\}^n$ with $|x| \leq k$, such that $a = B x$. That is, $B e = B x$ and the protocol behaves the same way on $e$ and $x$, which is a contradiction, since $f_k(e) = 1$ and $f_k(x)=0$.
\end{proof}

\begin{remark}
    Since $f_k$ has large (adaptive) parity decision tree complexity and for any $F : \{0, 1\}^n \times \{0, 1\}^n \to \{0, 1\}$ we have $\CCone(F) \geq \CC(F)$, all separations provided by functions $f_k$ translate into the same separations between $\DTP$ and $\CC$.
\end{remark}

Next, we proceed to the upper bound on the $\CCone(F_k)$.

\begin{theorem} \label{thm:separation-cc-to-covering}
    Suppose for some $n$, $k$ and $t$ there is a $(n, 2^t, R)$ covering code $\mathcal{C}$ for $R = \left\lfloor\frac{n-k-1}{2} \right \rfloor$. Then, $\CCone(F_k) \leq t$.
\end{theorem}

\begin{proof}
    Split the points of $\{0,1\}^n$ into balls with radius $R$ with centers in the points of $\mathcal{C}$ (if some point belongs to several balls, attribute it to one of them arbitrarily). This results in a partition of the cube into $2^t$ subsets with the diameter of each subset at most $n-k-1$. 

    The proof can be finished through Theorem~\ref{thm:ccone_to_graph_reduction}, but to make it more self-contained we directly describe communication protocol.

    On input $x$ Alice sends as $h(x)$ the index of the ball containing $x$. Bob computes $\neg y$, componentwise negation of $y$, and outputs 1 if it is in the same ball. If this is not the case, Bob outputs 0.

    Clearly, the complexity of this protocol is at most $t$. For the correctness of the protocol, if $f(x \oplus y) =1$, then $x = \neg y$ and the protocol clearly outputs $1$. However, if $f(x \oplus y) = 0$, then $|x \oplus y| \leq k$ and thus $\dist(x, \neg y) \geq n-k$. In this case $x$ and $\neg y$ are not in the same ball and the protocol outputs $0$.
\end{proof}

\begin{theorem} \label{thm:cc-upper-bound-through-codes}
    For any $n$ and $k$ we have 
    \begin{equation}
    \CCone(F_k) \leq n - \log V(n,R) + \log n
    \end{equation}
    for $R = \left\lfloor\frac{n-k-1}{2} \right \rfloor$.
\end{theorem}

\begin{proof}
    By Theorem~\ref{thm:covering-codes} there exists a $(n,2^t,R)$ covering code for
    \begin{equation}
    \log 2^t = t \leq n - \log V(n,R) + \log n.
    \end{equation}
    The theorem follows from Theorem~\ref{thm:separation-cc-to-covering}. 
\end{proof}

From this we can get a separation for a wide range of parameters. 

\begin{corollary}
    Suppose $k = cn$ for some constant $0 < c < 1$. Then $\NADTP(f_k) = cn+1$ and 
    \begin{equation}
    \CCone(F_k) \leq \left(1 - H\left(\frac{1-c}{2}\right)\right) n + O(\log n).
    \end{equation}
    In particular, $\CCone(F_k) < \NADTP(f_k)$.
    The number of undefined inputs for $f_k$ is $2^n - O(\frac{2^{H(c)n}}{\sqrt{n}})$ if $c< 1/2$, is equal to $(1+o(1))2^{n-1}$ if $c=1/2$, and is $O(\frac{2^{H(1-c)n}}{\sqrt{n}})$ if $c > 1/2$.
\end{corollary}

\begin{proof}
    The equality for $\NADTP$ is proved in Theorem~\ref{thm:counterexample_dtp_large}.

    For communication complexity bound we apply Theorem~\ref{thm:cc-upper-bound-through-codes}. We have $R = \left\lfloor\frac{(1-c)n-1}{2} \right \rfloor = \frac{(1-c)n}{2} + O(1)$ and by Lemmas~\ref{lem:binomial-linear-case} and~\ref{lem:entropy-convergence} we have
    \begin{equation}
    \log V(n,R) = H\left(\frac{1-c}{2}\right) n - O(\log n).
    \end{equation}
    By Theorem~\ref{thm:cc-upper-bound-through-codes} we have
    \begin{equation}
    \CCone(F_k) \leq n - \log V(n,R) + \log n = \left(1 - H\left(\frac{1-c}{2}\right)\right) n + O(\log n).
    \end{equation}
    To show that $\CCone(F_k) < \NADTP(f_k)$ we need to compare $k = cn$ with the bound on communication complexity. It is easy to see that
    \begin{equation}
    1 - H\left(\frac{1-c}{2}\right) < c
    \end{equation}
    for all $0<c<1$ (the left hand-side and the right hand-side are equal for $c=0$ and $c=1$ and the left hand-side is concave in $c$). 

    The bounds on the number of undefined inputs follow easily from Lemma~\ref{lem:binomial-linear-case}.
\end{proof}

The largest gap we can get is the following.

\begin{corollary}
    For $k = \Theta(\sqrt{n \log n})$ we have that $\NADTP(f_k) = \Theta(\sqrt{n \log n})$ and $\CCone(F_k) = O(\log n)$. The number of undefined inputs for $f_k$  is $2^n - 2^{\Theta(\sqrt{n} \log^{3/2} n)}$.
\end{corollary}

\begin{proof}
    For $k = \Theta(\sqrt{n \log n})$ we have $R = \frac n2 - \Theta(\sqrt{n \log n})$ in Theorem~\ref{thm:cc-upper-bound-through-codes}. By Lemma~\ref{lem:spencer} we have $V(n,R) = \frac{2^n}{\poly(n)}$ and as a result $\CCone(F_k) = O(\log n)$.

    For the number of undefined inputs, we apply Lemma~\ref{lem:binomial-universal}: 
    \begin{equation}
    \left( \frac{n}{k}\right)^k \leq  V(n,k) \leq \left( \frac{en}{k}\right)^k.
    \end{equation}
    For $k = \Theta(\sqrt{n \log n})$ it is easy to see that both sides are $2^{\Theta(\sqrt{n} \log^{3/2} n)}$. From this the estimate on the number of undefined inputs follows.
\end{proof}

\section{Extreme Cases}
\label{sec:extreme cases}

In this section we discuss extreme cases. All proves are moved to Appendix~\ref{sec:small_value_of_communication_complexity}.

For small values of complexity measures we have the following equality results.
\begin{theorem}
    \label{th_cut}
    Suppose $F$ satisfies $\CCone(F) = 1$. It then follows that $\NADTP(f) = 1$.
\end{theorem}

\begin{theorem} \label{thm:cc=2}
If function $f$ is undefined on fewer than $2^{n - 3}$ inputs and $\CCone(F) = 2$, then $\NADTP(f) = 2$.
\end{theorem}

On the other end of the spectrum, we show that if $\NADTP(f)$ is really large, then it is equal for all partial functions.

\begin{theorem} \label{thm:large-values}
    For any partial function $f \colon \{0,1\}^n \to \{0,1, \perp\}$, if $\NADTP(f) \geq n-1$, then $\CCone(F) = \NADTP(f)$.
\end{theorem}

The largest value of $\NADTP$ for which we get separation is $n-2$. 

\begin{theorem} \label{thm:large-separation}
    $\CCone(F_{n-3}) \leq n - \Theta(\log n)$, whereas $\NADTP(f_{n-3}) = n-2$. The number of undefined inputs for $f_{n-3}$ is $\frac{n(n+1)}{2}$.
\end{theorem}

The smallest value of $\CCone$ for which we get a separation is $7$. 

\begin{theorem} \label{thm:small-seration}
    For any $n \geq 32$ we have $\CCone(F_{7}) \leq 7$, whereas $\NADTP(f_{7}) = 8$.  
\end{theorem}


\bibliography{bibl}

\appendix

\section{Proofs for the Extreme Cases}
\label{sec:small_value_of_communication_complexity}
\subsection{Case \texorpdfstring{$\CCone(F) = 1$}{owCC=1}}
Assume there exists $t$-bit communication protocol, $(h, \varphi)$ for a function $F(x, y) = f(x \oplus y)$ where $f \colon \{0,1\}^n \to \{0,1\}$. A function $h$ is called \emph{balanced} if all classes in the $h$-induced partition are of equal size. We say that $h$ is \emph{balanced on a subset} when its restriction to the inputs in this subset is balanced. For the case $t = 1$, we analyze two distinct scenarios separately: when $h$ is balanced and when it is not. 

For the scenario where $h$ is unbalanced, we will demonstrate that all shifts are good, leading to the conclusion that $f$ is a constant function. Conversely, when $h$ is balanced, we identify a specific $n - 1$-dimensional subspace of $\{0, 1\}^n$ on which $h$ is unbalanced. We then show that every shift in this subspace is good. This observation gives us that the function value of $f$ depends solely on whether $x$ belongs to this identified subspace and this can be checked with a single parity query.

\begin{lemma}
    \label{lemma:unbalanced_function_with_ccone_1}
    Assume there exists $1$-bit communication protocol, $(h, \varphi)$ for a function $F(x, y) = f(x \oplus y)$ where $f \colon \{0,1\}^n \to \{0,1\}$. If $h$ is unbalanced, then every shift is good.
\end{lemma}
\begin{proof}
    Consider arbitrary shift $\Delta$. Consider the cosets corresponding to the subspace $\langle \Delta \rangle$. The $h$-induced partition consists of two classes, since they are not equal, one class contains more than $2^{n - 1}$ elements. Applying the Pigeonhole principle we get that some coset of the subspace $\langle \Delta \rangle$ contains two elements with the same $h$ value. Given that a coset has only two points and those differ by shift $\Delta$, we conclude that $\Delta$ is indeed a good shift.
\end{proof}

\begin{lemma}
    \label{lemma:unbalanced_on_subspace_function}
    Assume there exists $1$-bit communication protocol, $(h, \varphi)$ for a function $F(x, y) = f(x \oplus y)$ where $f \colon \{0,1\}^n \to \{0,1\}$. If $h$ is unbalanced on a given subspace of $\{0, 1\}^n$, then every shift in this subspace is good.
\end{lemma}
\begin{proof}
    The proof is completely analogous to the proof of Lemma~\ref{lemma:unbalanced_function_with_ccone_1}. Indeed, since $h$ is unbalanced on the subspace, for any shift $\Delta$ in the subspace there are $x$ and $y$ such that $h(x)=h(y)$ and $x \oplus y = \Delta$. Thus, $\Delta$ is a good shift.
\end{proof}

\begin{lemma}
    \label{lemma:can_select_unbalanced_subspace}
    Suppose there exists $t$-bit communication protocol, $(h, \varphi)$ for a function $F(x, y) = f(x \oplus y)$ where $f \colon \{0,1\}^n \to \{0,1\}$. For a balanced function $h$, there is an $n - 1$-dimensional subspace of of $\{0, 1\}^n$ over which $h$ is unbalanced.
\end{lemma}
\begin{proof}
    The proof is based on Fourier analysis. For the completeness of the proof, we provide basic definitions in Appendix~\ref{sec:fourier}.
    
    Consider Fourier decomposition of $h$. Since $h$ is balanced and thus not constant, there must be a non-zero coefficient $\hat h(S)$ in its Fourier decomposition associated with a non-empty subset $S$. We show that $h$ is unbalanced on the $n-1$-dimensional linear subspace $X = \{x | \chi_{S}(x) = 1\}$. Assume, for the sake of contradiction, that $h$ is balanced on $X$. 
    The Fourier coefficient $\hat{h}(S)$ can be computed as follows:
    \begin{multline}
    \hat{h}(S) = \frac{1}{2^n}\sum_{x}(-1)^{h(x)} \chi_S(x) =\\ 
    \frac{1}{2^n}\Big(|\{h(x) = 0, x \in X\}| - |\{h(x) = 1, x \in X\}| - |\{h(x) = 0, x \notin X\}| + |\{h(x) = 1, x \notin X\}|\Big).
    \end{multline}
    Denote the quantity $|\{h(x) = 0, x \in X\}|$ as $a$. As $h$ is balanced on $X$, it follows that $|\{h(x) = 1, x \in X\}| = a$. The set $X$ contains $2^{n-1}$ elements so $a = 2^{n - 2}$. Given that $h$ is balanced across $\{0, 1\}^n$, both the sets $\{h(x) = 0, x \in \{0, 1\}^n\}$ and $\{h(x) = 1, x \in \{0, 1\}^n\}$ each have $2^{n - 1}$ elements. Therefore:
    \begin{align}
    |\{h(x) = 0, x \notin X\}| = |\{h(x) = 0, x \in \{0, 1\}^n\}| - |\{h(x) = 0, x \in X\}| = 2^{n - 2},\\
    |\{h(x) = 1, x \notin X\}| = |\{h(x) = 1, x \in \{0, 1\}^n\}| - |\{h(x) = 1, x \in X\}| = 2^{n - 2}.
    \end{align}
    We can see that $\hat h(S) = 0$ which leads us to the required contradiction.
\end{proof}

We are now ready to prove Theorem~\ref{th_cut}

\begin{proof}[Proof of Theorem~\ref{th_cut}]
        Given $1$-bit communication protocol, $(h, \varphi)$ for a function $F$, consider the total $h$-induced graph. For any unbalanced $h$ by Lemma~\ref{lemma:unbalanced_function_with_ccone_1} we get that all shifts are good, so the graph is complete. It can't be split into connectivity components by vertex removal, therefore the partial $h$-induced graph has a single connectivity component. By Corollary~\ref{cor:bounding_nadt_through_ccone} we have $\NADTP(f) = 0$.
        
        For a balanced function, we use Lemma~\ref{lemma:can_select_unbalanced_subspace} to choose an $n - 1$-dimensional subspace $U$ of $\{0, 1\}^n$, on which $h$ is unbalanced. By Lemma~\ref{lemma:unbalanced_on_subspace_function}, all the shifts in $U$ are good. Select two arbitrary vertices $x$ and $y$, from the same coset of $U$. Vertices $x$ and $y$ are connected in the total $h$-induced graph because their XOR belongs to $U$. Therefore cosets of $U$ are cliques and they will remain connected in a partial $h$-induced graph. By Corollary~\ref{cor:bounding_nadt_through_ccone} we conclude that $\NADTP(f) = 1$.
\end{proof}

\label{sec_one}
\subsection{Case \texorpdfstring{$\CCone(F) = 2$}{owCC=2}}
\label{sec_two}
Assume there exists $2$-bit communication protocol, $(h, \varphi)$ for a function $F(x, y) = f(x \oplus y)$ where $f \colon \{0,1\}^n \to \{0,1\}$.
We handle cases when $h$ is unbalanced and balanced separately. In the first case, we observe that the XOR of two bad shifts results in a good shift. We then use a known result on the bound on sumset cardinality to show that the good shifts either contain a coset of a $n - 1$-dimensional subspace of $\{0, 1\}^n$ or there exists \emph{large enough} number of such shifts. Either of these cases implies a certain structure on the total $h$-induced graph, which allows us to get the desired lower bound. When $h$ is balanced, we again consider the subspace of $\{0, 1\}^n$ on which it is unbalanced and analogously to the prior scenario, we deduce a specific structure on the subspace allowing us to conclude the proof.
\begin{lemma}
    \label{lemma:sum_of_bad_shifts_is_good}
    Assume there exists $2$-bit communication protocol, $(h, \varphi)$ for a function $F(x, y) = f(x \oplus y)$ where $f \colon \{0,1\}^n \to \{0,1\}$ and the function $h$ is unbalanced. Then the XOR of two bad shifts is a good shift.
\end{lemma}
\begin{proof}
    Assume $\Delta_1$ and $\Delta_2$ are bad shifts. Consider the cosets of the subspace $\langle \Delta_1, \Delta_2 \rangle$. There are a total of $2^{n - 2}$ such cosets. As the function $h$ is unbalanced, the $h$-induced partition has a class, denoted as $H_1$, which contains strictly more than $2^{n - 2}$ elements. By the Pigeonhole principle, there exists a coset of $\langle \Delta_1, \Delta_2 \rangle$ that contains two elements, namely $x$ and $y$, both of which belong to $H_1$. As $h(x) = h(y)$, the XOR of $x$ and $y$ produces a good shift. Additionally, $x$ and $y$ lay in the same coset, thus the shift $x \oplus y$ is a member of $\langle \Delta_1, \Delta_2 \rangle$. Within the subspace $\langle \Delta_1, \Delta_2 \rangle$, there are only three distinct non-zero shifts: $\Delta_1$, $\Delta_2$, and $\Delta_1 \oplus \Delta_2$. Given that both $\Delta_1$ and $\Delta_2$ are bad shifts, the only possible good shift among them is $\Delta_1 \oplus \Delta_2$.
\end{proof}

\begin{theorem}
\label{thm:sumset_bound}
  Let $A$ and $B$ be non-empty subsets of $\{0, 1\}^n$. Define the sumset of $A$ and $B$ as $A + B = \{a + b | a \in A, b \in B\}$. Assume that $A$ is not contained in a coset of any proper subspace of $\{0, 1\}^n$. Then
  \begin{equation}
  |A + B| \geq \min\{|A| + |B| - 2^{n-3}, 3 \cdot 2^{n - 2}\}.
  \end{equation}
\end{theorem}
The proof of this theorem is moved to Appendix~\ref{app:set-sum}.
\begin{lemma}
\label{lemma:either_good_shift_make_subspace_or_there_are_many_of_those}
    Assume there exists $2$-bit communication protocol, $(h, \varphi)$ for a function $F(x, y) = f(x \oplus y)$ where $f \colon \{0,1\}^n \to \{0,1\}$ and $h$ is unbalanced. Then either there exists at least $5 \cdot 2^{n - 3} - 1$ good shifts (not counting 0), or the set of good shifts contains a coset of an $n-1$-dimensional subspace of $\{0, 1\}^n$.
\end{lemma}
\begin{proof}
        Let $B$ be the set of bad shifts and $\overline{B}$ be the set of good shifts, these are complementary so $|B| + |\overline{B}| = 2^n$. There are two cases to consider: either $B$ is a subset of a coset of a proper subspace of $\{0, 1\}^n$ or it is not.
        In the first case, let $Q$ be a subspace of $\{0, 1\}^n$ and $q$ be a vector in $\{0, 1\}^n$ such that $B \subseteq Q + q$. We extend the coset $Q + q$ to a coset $Q' + q$ of some $n - 1$-dimensional subspace $Q'$ of $\{0, 1\}^n$. Observe that since $B$ is fully contained in $Q' + q$, another coset of $Q'$ it is fully contained in $\overline{B}$.
        
        In the second case, first observe that by Lemma~\ref{lemma:sum_of_bad_shifts_is_good} the sum of bad shifts is a good shift, thus we have $B + B \subseteq \overline{B}$. By Theorem~\ref{thm:sumset_bound} we have
        \begin{equation}
        |\overline{B}| \geq |B + B| \geq \min\{2|B| - 2^{n-3}, 3 \cdot 2^{n - 2}\}.
        \end{equation}
        We also know that $|B| + |\overline{B}| = 2^n$. As a result, either
        \begin{equation}
        |B| + 2|B| - 2^{n-3} \leq 2^n,
        \end{equation}   
        or
        \begin{equation}
        |\overline{B}| \geq 3 \cdot 2^{n - 2}.
        \end{equation}   
        It is easy to see that in both cases
        \begin{equation}     
        |\overline{B}| \geq 5 \cdot 2^{n - 3}.
        \end{equation}     
        If we exclude the zero shift, we have at least $5 \cdot 2^{n - 3} - 1$ good shifts.
\end{proof}

\begin{lemma}
\label{lemma:unbalanced_function_with_ccone_2}
Assume there exists $2$-bit communication protocol, $(h, \varphi)$ for a function $F(x, y) = f(x \oplus y)$ where $f \colon \{0,1\}^n \to \{0,1\}$. If $h$ is unbalanced, then one of the following two conditions is true for the total $h$-induced graph:
\begin{itemize}
\item Total $h$-induced graph consists of two cliques, each being a coset of an $n-1$-dimensional subspace of $\{0, 1\}^n$.
\item Total $h$-induced graph is $2^{n-2}$-vertex connected.
\end{itemize}
\end{lemma}
\begin{proof}
We consider three cases.

\textbf{Case 1:} In this case, we assume that the set of good shifts contains a $n-1$-dimensional subspace $Q$ of $\{0, 1\}^n$. Take two arbitrary points $x$ and $y$ from the same coset $Q + q$, where $q$ is a specific vector in $ \{0, 1\}^n $. Then, $x$ and $y$ can be expressed as $x = x' \oplus q$ and $y = y' \oplus q$ for $x', y' \in Q$. Consequently, $x \oplus y = x' \oplus y' \in Q$. This shows that any two points in the coset of $Q$ are connected by an edge in the total $h$-induced graph, forming cliques.

\textbf{Case 2:} Assume that the set of good shifts contains an $n-1$-dimensional coset $Q + q$, where $Q$ is an $n-1$-dimensional subspace of $\{0, 1\}^n$ and $q$ is a vector not in $Q$. Consider two arbitrary points $x$ and $y$ from different cosets of $Q$. Without loss of generality, let $x \in Q$ and $y \in Q + q$. There exists $y' \in Q$ such that $y = y' \oplus q$. Then, $x \oplus y = (x \oplus y') \oplus q \in Q + q$. Thus, an edge exists between $x$ and $y$ in the total $h$-induced graph, and, as a result, the graph contains a complete bipartite graph with parts being the cosets of $Q$. To make this graph disconnected one has to delete the whole part, thus the graph is $2^{n - 1}$-connected.

\textbf{Case 3:} Assume the set of good shifts satisfies neither of the first two conditions. Then, by Lemma~\ref{lemma:either_good_shift_make_subspace_or_there_are_many_of_those}, there must be at least $5 \cdot 2^{n-3} - 1$ good shifts. Take any two arbitrary non-neighboring vertices $x$ and $y$; the sizes of their neighbor sets are at least $5 \cdot 2^{n-3} - 1$. Given that the total number of vertices excluding $x$ and $y$ is $2^n - 2$, the intersection of these neighbor sets must contain at least $2^{n-2}$ vertices. Hence, removing fewer than $2^{n-2}$ vertices cannot disconnect the graph.
\end{proof}

\begin{lemma}
\label{lemma:unbalanced_on_subspace_function_with_ccone_2}
Assume there exists $2$-bit communication protocol, $(h, \varphi)$ for a function $F(x, y) = f(x \oplus y)$ where $f \colon \{0,1\}^n \to \{0,1\}$. If $h$ is unbalanced on an $n-1$-dimensional subspace $Q$ of $\{0, 1\}^n$, then one of the following conditions must hold:
\begin{itemize}
\item The total $h$-induced graph consists of four distinct cliques, each of which corresponds to a coset of an $n-2$-dimensional subspace of $Q$.
\item The subgraphs of the total $h$-induced graph on the vertices of cosets of $Q$, are at least $2^{n - 3}$-vertex connected.
\end{itemize}
\end{lemma}
\begin{proof}
For the proof we just apply Lemma~\ref{lemma:unbalanced_function_with_ccone_2} on the subspace $Q$. Formally,
let $B$ be a matrix whose columns form a basis for $Q$. We  define a new function $h': x \mapsto h(Bx)$ ($x$ is of length $n-1$).
Applying Lemma~\ref{lemma:unbalanced_function_with_ccone_2}, we conclude that the total $h'$-induced graph either consists of cliques corresponding to cosets of an $n-2$-dimensional subspace $Q'$ of $\{0, 1\}^n$ or that graph is $2^{n - 3}$-vertex connected.

To relate $h'$ back to $h$, we consider a vector $q$ not in $Q$ and define two graph embeddings $\psi_1: x \mapsto Bx$ and $\psi_2: x \mapsto Bx \oplus q$ of the total $h'$-induced graph into the total $h$-induced graph. The images of these mappings are $Q$ and $Q + q$. To see that they are indeed graph embeddings we notice that if $x$ and $y$ are connected in the total $h'$-induced graph, $x \oplus y$ is a good shift for $h'$, so $B(x \oplus y)$ is a good shift for $h$, which implies that images of $x$ and $y$ under $\psi_1$, as well as images of $x$ and $y$ under $\psi_2$, are indeed connected in $h$-induced graph. The bound on vertex connectivity of cosets follows from these embeddings. Note that these mappings are also affine transformations that only differ by a shift. Therefore, the image of cosets in $\{0, 1\}^{n - 1}$ over these mappings will result in cosets of the same space in $\{0, 1\}^n$, which finishes the proof.
\end{proof}

We are now ready to finish the proof of Theorem~\ref{thm:cc=2}.

\begin{proof}[Proof of Theorem~\ref{thm:cc=2}]
Given $2$-bit communication protocol, $(h, \varphi)$ for a function $F$,
we have two main cases to consider, depending on whether $h$ is balanced or unbalanced. If $h$ is unbalanced, we apply Lemma~\ref{lemma:unbalanced_function_with_ccone_2}. As a result, either the $h$-induced graph consists of cliques corresponding to cosets of $n-1$-dimensional subspace of $\{0, 1\}^n$, or the $h$-induced graph is $2^{n-2}$-vertex connected. In the first case, by Corollary~\ref{cor:bounding_nadt_through_ccone}, we conclude that $\NADTP(f) \leq 1$, which is a contradiction.
In the second case the graph is $2^{n-2}$-vertex connected and again by Corollary~\ref{cor:bounding_nadt_through_ccone} we find that $\NADTP(f) = 0$ because the function $f$ is undefined on fewer than $2^{n-2}$ inputs, making it impossible to disconnect the graph by removing these vertices.

If $h$ is balanced, we use Lemma~\ref{lemma:can_select_unbalanced_subspace} to find a subspace $Q$ of $\{0, 1\}^n$ where $h$ becomes unbalanced. Then by Lemma~\ref{lemma:unbalanced_on_subspace_function_with_ccone_2} the graph will split either into four fully connected cosets of subspace of $\{0, 1\}^n$, or into two $2^{n - 3}$ vertex-connected cosets of subspace of $\{0, 1\}^n$. As $f$ in undefined in less than $2^{n - 3}$ points we again use Corollary~\ref{cor:bounding_nadt_through_ccone} and conclude that $\NADTP(f) \leq 2$. 
\end{proof}

\subsection{Equality for Large Values of \texorpdfstring{$\CCone(F)$}{owCC} and \texorpdfstring{$\NADTP(f)$}{PDT}}

\begin{proof}[Proof of Theorem~\ref{thm:large-values}]
    First consider the case $\NADTP(f) = n$ and assume that $\CCone(F) \leq n-1$. Consider the corresponding function $h$. One of its equivalence classes $H$ is of size at least $2$, denote two of its elements by $u$ and $v$. We have that $\Delta = u \oplus v$ is a good shift. Thus, for any $x$ if $f(x)$ and $f(x \oplus \Delta)$ are defined, then $f(x) = f(x \oplus \Delta)$. But this exactly means that there is a 1-dimensional space such that $f$ is constant on each of its cosets. Thus, $\NADTP(f) \leq n-1$, which is a contradiction.

    Now consider the case $\NADTP(f) = n-1$ and again assume that $\CCone(F) \leq n-2$. Consider the corresponding function $h$. Now one of its equivalence classes $H$ is of size at least $4$. Consider any three points $u$, $v$, $w$ in this class. Then the vectors $u \oplus v$, $u \oplus w$ and $v \oplus w$ are good shifts. Note that they together with 0-vector form a 2-dimensional linear subspace of $\{0, 1\}^n$ consisting of good shifts. As a result, $f$ is a constant on every coset of this subspace and $\NADTP(f) \leq n-2$, which is a contradiction.
\end{proof}

\subsection{Separations in Boundary Cases}
\label{sec:sep-boundary}

\begin{proof}[Proof of Theorem~\ref{thm:large-separation}]
    We have already proved equality for $\NADTP(f_{n-3})$ and it remains to bound $\CCone(F_{n-3})$.

    For this we use Theorem~\ref{thm:separation-cc-to-covering}. Note that in our case $R = \lfloor\frac{n-k-1}{2} \rfloor = 1$.

    If $n = 2^m - 1$ for some integer $m$, then we can just use Theorem~\ref{thm:hamming-codes}. Each ball of radius 1 is of volume $n+1$ and thus in total we have $2^n/(n+1)$ balls. As a result, 
    \begin{equation}
    \CCone(F_{n-3}) \leq \log \frac{2^n}{n+1} = n - \log (n+1).
    \end{equation}

    For general $n$ consider maximal integer $m$ such that $2^m - 1 \leq n$. Denote  $n_1 = 2^m - 1$ and $n_2 = n - n_1$. Consider Hamming code $\mathcal{C}_1$ on $\{0,1\}^{n_1}$ and consider the code $\mathcal{C}_2 = \{0,1\}^{n_2}$. The latter code has parameters $(n_2, 2^{n_2}, 0)$. By Lemma~\ref{lem:direct-sum} we have that $\mathcal{C}_1 \oplus \mathcal{C}_2$ has parameters $(n, \frac{2^{n_1}}{n_1 + 1} \cdot 2^{n_2}, 1)$. Since $n_1$ is at least half of $n$, we have 
    \begin{equation}
    \CCone(F_{n-3}) \leq \log \left(\frac{2^{n_1}}{n_1 + 1} \cdot 2^{n_2} \right)  = n - \Theta(\log n).
    \end{equation}

    The undefined inputs of $f_{n-3}$ are just inputs $x \in \{0,1\}^n$ with weight $n-1$ and $n-2$. It is easy to see that there are $\frac{n(n+1)}{2}$ of them.    
\end{proof}

\begin{proof}[Proof of Theorem~\ref{thm:small-seration}]
    Again, we already found $\NADTP(f_7)$.

    For the bound on $\CCone$ we start with Reed-Muller code $\mathcal{RM}(1,5)$~\cite[Chapter 9]{CohenHLL97}. This code has parameters $(2^5, 2^6, 12)$ (as a covering code), that is, it has $32$ input bits, the number of covering balls is $2^6$ and their radius is $R = 12$. In terms of Theorem~\ref{thm:separation-cc-to-covering} we have  $R = \frac{32 - 7 - 1}{2}$ and thus the code gives us the protocol for $F_7$ of size $\log 2^6 = 6$ on $n=32$ inputs (that is, for the particular case of $n=32$ we have an even better upper bound on communication complexity).

    For general $n \geq 32$ denote $n_1 = 32$ and $n_2 = n - n_1$. Let $\mathcal{C}_1$ be Reed-Muller code introduced above and $\mathcal{C}_2$ consist of two vectors: all zeros and all ones. The code $\mathcal{C}_2$ has parameters $(n_2, 2, \left \lfloor \frac{n_2}{2} \right \rfloor)$. Then $\mathcal{C}_1 \oplus \mathcal{C}_2$ has parameters $(n, 2^7, \left \lfloor \frac{n_2}{2} \right \rfloor + 12)$. Note that its radius $R$ can be bounded as
    \begin{equation}
    R =\left \lfloor \frac{n_2}{2} \right \rfloor + 12 \leq  \frac{n_2}{2}  + 12  = \frac n2 + 12 - \frac{32}{2} = \frac{n - 7 - 1}{2}.
    \end{equation}
    Thus, the code gives a protocol for $F_7$ of size $7$.
\end{proof}

\section{Isoperimetric Inequalities}
\label{app:isoperim}
This section is devoted to proving Lemma~\ref{lemma:isoperim_final}.
\begin{lemma}
    \label{lemma:isoperim_border}
    For any subset $A \subseteq \{0, 1\}^m$ of $m$-dimensional Boolean cube vertices, it holds that $|\Gamma' A| \geq |\Gamma' I^m_{|A|}|$. 
\end{lemma}
\begin{proof}
    In the case $|A| = 1$, $A$ and $I^M_{|A|}$ are just sets of single element and equality between $|\Gamma' A|$ and $|\Gamma' I^m_{|A|}|$ is obvious. Otherwise the set $I^m_{|A|}$ doesn't have isolated vertices. Thus, all the vertices in $I^m_{|A|}$ are neighbors of $I^m_{|A|}$ and $|\Gamma' I^m_{|A|}| = |\Gamma I^m_{|A|}| - |A|$. Meanwhile $|\Gamma' A| \geq |\Gamma A| - |A|$. Therefore Theorem~\ref{thm:harper} implies
    \begin{equation}
    |\Gamma' A| \geq |\Gamma A| - |A| \geq |\Gamma I^m_{|A|}| - |A| = |\Gamma' I^m_{|A|}|.
    \end{equation}
\end{proof}
\begin{lemma}
    \label{lemma:isoperim_monotone_size}
    For $a$ satisfying $V(m, r) \leq a \leq V\left(m, \left\lfloor \frac{m - 1}{2}\right\rfloor\right)$
    the following holds: 
    \begin{equation}
    |\Gamma' I^{m}_{a}| \geq |\Gamma' I^{m}_{V(m, r)}| = \binom{m}{r + 1}.\end{equation}
\end{lemma}
\begin{proof}
        Let $r'$ be the maximum integer for which $V(m, r') \leq a$. Note that $r' \geq r$. If $a = V(m, r')$, the lemma is trivial. Otherwise, the inequality $a \leq V\left(\left\lfloor \frac{m - 1}{2}\right\rfloor, m\right)$ implies that $r' \leq \left\lfloor \frac{m - 1}{2}\right\rfloor - 1$. 

        The set $I^{m}_{a}$ contains elements with Hamming weight up to $r'$ and possibly some with weight $r' + 1$. Let $B = I^{m}_{a} \setminus I^{m}_{V(m, r')}$ be the elements of $I^m_{a}$ with Hamming weight $r' + 1$. Define 
        \begin{equation}
        B^{+} = \{x \in \Gamma' B: |x| = r' + 2\}.
        \end{equation}
        Elements of $B$ doesn't belong to $\Gamma' I^m_{a}$, since they belong to $I^m_{a}$, meanwhile elements of $B^{+}$ belong to $\Gamma' I^m_{a}$, since they are neighbors of elements from $B$ and doesn't belong $I^m_{V(m, r')}$. Therefore,
        \begin{equation}
        |\Gamma' I^m_{a}| = \binom{m}{r' + 1} - |B| + |B^{+}|.
        \end{equation}
        To prove that $|B^{+}| \geq |B|$, let's consider a bipartite subgraph $G$ of $m$-dimensional Boolean cube. The left part contains vertices with Hamming weight $r' + 1$, and the right part contains vertices with Hamming weight $r' + 2$. Here, $B$ is a subset of the left part, and $B^{+}$ is the set of neighbors of $B$ in $G$. Note that the degree of any vertex in the left part is 
        \begin{equation}
        \deg_L = m - (r' + 1) \geq m - \lfloor \frac{m - 1}{2}\rfloor = \lceil\frac{m - 1}{2}\rceil + 1,
        \end{equation}
        while the degree of any vertex in the right part is 
        \begin{equation}
        \deg_R = r' + 2 \leq \lfloor\frac{m - 1}{2}\rfloor + 1.
        \end{equation}
        Given that edges from $B$ connect exclusively to vertices in $B^{+}$, it follows that $|B|\deg_L \leq |B^{+}|\deg_R$, which implies $|B^{+}| \geq |B|$. Consequently, 
        \begin{equation}
        |\Gamma' I^m_{a}| \geq |I^m_{V(m, r')}| = \binom{m}{r' + 1} \geq \binom{m}{r + 1}.
        \end{equation}
\end{proof}
\begin{remark}
\label{remark:isoperim_lower_bound}
    A similar idea applies for $a$ larger then $V\left(m, \left\lfloor \frac{m - 1}{2}\right\rfloor\right)$. In that case $|B^{+}| \geq \frac{\deg_L}{\deg_R}|B| = \frac{m - (r' + 1)}{r' + 2}|B|$, therefore $|\Gamma' I^{m}_{a}| \geq \frac{m - (r' + 1)}{r' + 2}|\Gamma' I^{m}_{V(m, r')}|$. Note that here, unlike in previous case, $r'$ must be the largest integer satisfying $V(m, r') \leq a$.
\end{remark}

\begin{lemma}
\label{lemma:isoperim_monotone_dimension}
For any $M \geq m$ and any $a \leq 2^m$ it holds that $|\Gamma' I^m_a| \leq |\Gamma' I^{M}_a|$.
\end{lemma}
\begin{proof}
The proof goes by induction on $M$. The base case for $M = m$ is trivial. Assuming the lemma holds for $M$, we aim to prove it for $M + 1$. For this we construct a subset $A \subseteq \{0, 1\}^M$ with $|A| = a$ and $|\Gamma' A| \leq |\Gamma' I_{a}^{M + 1}|.$
Here, the first $\Gamma'$ refers to the $M$-dimensional Boolean cube, while the second $\Gamma'$ refers to the $(M + 1)$-dimensional Boolean cube.

We consider the 'slices' of the set $I^{M + 1}_{a}$ along its last coordinate:
\begin{align}
A_0 = \{(x_1, \ldots, x_{M}): x \in I^{M + 1}_{a}, x_{M + 1} = 0\},\\
A_1 = \{(x_1, \ldots, x_{M}): x \in I^{M + 1}_{a}, x_{M + 1} = 1\}.
\end{align}
Denote by $r$ the maximum number such that all the elements with Hamming weight at most $r$ belong to $I_{a}^{M + 1}$. The set $A_0$ contains all the elements with Hamming weight $r$ and maybe some elements with Hamming weight $r + 1$, while the set $A_1$ contains all the elements with Hamming weight $r - 1$ and maybe some elements with Hamming weight $r$. Three cases arise based on the dimension $M$: either $2r + 1 < M$, $2r + 1 = M$ or $2r=M$. As $a \leq 2^m \leq 2^M$ it's impossible for $r$ to take larger values. The third case is trivial, here $a$ is just equal to $2^M$ and boundary is empty. 

In the first case, we define 
\begin{equation}
A = A_0 \sqcup \lnot A_1,
\end{equation}
where 
\begin{equation}
\lnot A_1 = \{(1-x_1, \ldots, 1-x_{M}): x \in A_1\}.
\end{equation}
This union is indeed disjoint because the first set has elements with Hamming weight not above $r + 1$, while the second has elements with weight at most $M - r$. Next, we notice that the cardinality of the boundary of $A$ does not exceed that of $I^{M + 1}_{a}$. Indeed, if a vertex belongs to $\Gamma' A$ it either belongs to $\Gamma' A_0$ or to $\Gamma'  \lnot A_1$ or to both. That is, $|\Gamma' A| \leq |\Gamma' A_0| + |\Gamma' \lnot A_1|$. As we get $\lnot A_1$ from $A_1$ with graph automorphism, $|\Gamma' \lnot A_1| = |\Gamma' A_1|$. If vertex $v$ belongs $\Gamma' A_0$, then vertex $(v, 0)$ belongs $\Gamma' I_{a}^{m' + 1}$ and similarly if $v$ belongs $\Gamma' A_1$, then $(v, 1)$ belongs $\Gamma' I^{m' + 1}_a$. Therefore, 
\begin{equation}
|\Gamma' I^{m}_a| \leq |\Gamma' I^{M}_a| \leq |\Gamma' A| \leq |\Gamma' I^{M + 1}_a|.
\end{equation}

In the second case, we adjust the construction of $A$ because otherwise points from $A_0$ and $\lnot A_1$ may overlap. The set $A$ contains all vertices with Hamming weight at most $r$ and at least $M - r + 1$, and is filled up to cardinality $a$ with vertices having Hamming weight $r + 1 = M - r$. In this configuration, $\Gamma' A$ contains vertices of Hamming weight $r + 1$ that are not in $A$. But the number of such elements doesn't exceed the number of elements with weight $r + 1$, which doesn't belong to $A_0$ and all these elements lay in $\Gamma' A_0$, hence:
\begin{equation}
|\Gamma' I^{m}_a| \leq |\Gamma' I^{M}_a| \leq |\Gamma' A| \leq |\Gamma' A_0| \leq |\Gamma' I^{M + 1}_{a}|.
\end{equation}
This finishes the proof of the induction step and the lemma.
\end{proof}

\begin{lemma}
\label{lemma:isoperim_hamming_ball_of_larger_dim_near_half}
For all $M$ there exists such $r$ that 
\begin{equation}
V\left(M - 1, \left\lfloor\frac{M - 2}{2}\right\rfloor - 2\right) \leq V\left(M, r\right) \leq V\left(M - 1, \left\lfloor\frac{M - 2}{2}\right\rfloor\right).
\end{equation}
\end{lemma}
\begin{proof}
We select $r$ to be the smallest number such that $V\left(M - 1, \left\lfloor\frac{M - 2}{2}\right\rfloor - 2\right) \leq V\left(M, r\right)$. Clearly, $r \leq \left\lfloor\frac{M - 2}{2}\right\rfloor - 2$. For such $r$ the following holds:
\begin{equation}
V\left(M - 1, \left\lfloor\frac{M - 2}{2}\right\rfloor - 2\right) \leq V(M, r) \leq V\left(M - 1, \left\lfloor\frac{M - 2}{2}\right\rfloor - 2\right) + \binom{M}{r}.
\end{equation}
From here, we can further bound $\binom{M}{r}$ as follows:
\begin{equation}
\binom{M}{r} = \frac{M}{M - r}\binom{M - 1}{r} \leq 2\binom{M - 1}{r}.
\end{equation}
Thus,
\begin{equation}
V(M, r) \leq V\left(M - 1, \left\lfloor \frac{M - 2}{2}\right\rfloor - 2\right) + 2\binom{M - 1}{r} \leq V\left(M - 1, \left\lfloor \frac{M - 2}{2} \right\rfloor\right).
\end{equation}
The last inequality holds since $r \leq \left\lfloor \frac{M - 2}{2} \right\rfloor - 2$.
\end{proof}
\begin{proof}[Proof of Lemma~\ref{lemma:isoperim_final}]
    First, we consider the case when $|A| \leq V\left(k, \left\lfloor\frac{k - 1}{2}\right\rfloor\right)$. Here we let $M = k$ and $a = |A|$. By Lemma~\ref{lemma:isoperim_border} we have $|\Gamma' A| \geq |\Gamma' I_{a}^{M}|$. We will iteratively decrease $M$ and $a$ until $M = m$ and $a = V(m, \lfloor{\frac{m - 1}{2}}\rfloor - 2)$ in a way that the boundary of the set $I^{M}_{a}$ does not increase. When the algorithm finishes, the set $I^{M}_{a}$ is a Hamming ball and its boundary contains all the elements with weight $\left\lfloor\frac{m - 1}{2}\right\rfloor - 1$ and thus is of volume $\binom{m}{\lfloor{\frac{m - 1}{2}}\rfloor - 1}$. The size of the boundary of an initial set is at least as large. 
    
    We decrease the variables in the following way. While $M$ is larger then $m$, if $a \leq V(M - 1, \left\lfloor\frac{M - 2}{2}\right\rfloor)$ we simply apply Lemma~\ref{lemma:isoperim_monotone_dimension} to decrease $M$ by one, otherwise we first set $a$ to be $V(M, r)$, where $r$ is selected by Lemma~\ref{lemma:isoperim_hamming_ball_of_larger_dim_near_half}, the boundary won't increase after these assignment by Lemma~\ref{lemma:isoperim_monotone_size} and then we again apply Lemma~\ref{lemma:isoperim_monotone_dimension} to decrease $M$. On all steps of the algorithm, $a$ doesn't exceed $V(M, \left\lfloor\frac{M - 1}{2}\right\rfloor)$ which allows us to use these lemmas. When $M$ reaches $m$ it holds that $V(m, \left\lfloor\frac{m - 1}{2}\right\rfloor - 2) \leq a \leq V(m, \left\lfloor\frac{m - 1}{2}\right\rfloor)$ and we make $a$ to be precisely equal to $V(m, \left\lfloor\frac{m - 1}{2}\right\rfloor - 2)$ by applying Lemma~\ref{lemma:isoperim_monotone_size} once again. 
    
    There exists a remaining case if initially $V(k, \left\lfloor\frac{k - 1}{2}\right\rfloor) \leq |A| \leq 2^{k - 1}$. It is only possible if $k$ is even. In that case we use Remark~\ref{remark:isoperim_lower_bound} with $r' = \left\lfloor\frac{k - 1}{2}\right\rfloor$ to conclude that 
    \begin{equation}
    |\Gamma' A| \geq \frac{k}{k + 2}\binom{k}{\frac{k}{2}} = \binom{k}{\frac{k}{2} - 1} = |\Gamma' I^k_{V(k, k/2-2)}|.
    \end{equation}  
    As $\left(k, \left\lfloor\frac{k - 1}{2}\right\rfloor - 2\right) \leq V(k, \frac{k}{2} - 2) \leq V\left(k, \left\lfloor\frac{k - 1}{2}\right\rfloor\right)$ the statement of the lemma follows from the first case.
\end{proof}

\section{Fourier Analysis}
\label{sec:fourier}
Here we provide the basic definitions from Fourier analysis. Functions that map $\{0, 1\}^n \to \R$ form a $2^n$-dimensional vector space under the operation of addition (indeed we can represent the function as a $2^n$-dimensional vector of values for each of $n$-bit binary strings). For this space, we introduce an inner product:
\begin{equation}
\langle \psi, \theta\rangle = \frac{1}{2^n}\sum_{x \in \{0, 1\}^n} \psi(x) \theta(x).
\end{equation}
Let's consider the parity functions, which are expressed as
\begin{equation}
\chi_{S}(x) = (-1)^{\sum_{i \in S}x_i},
\end{equation}
with $S \subseteq [n]$. These functions form an orthonormal basis with respect to our previously defined inner product. As a direct consequence, any function $\psi$ of the form $\{0, 1\}^n \to \R$ can be uniquely represented as
\begin{equation}
\psi(x) = \sum_{S\subseteq [n]} \hat \psi(S) \chi_S(x).
\end{equation}
The terms $\hat \psi(S)$ in the above expansion are known as Fourier coefficients. They can be computed in the following way:
\begin{equation}
\hat \psi(S) = \langle \psi, \chi_S\rangle = \frac{1}{2^n}\sum_{x \in \{0, 1\}^n} \psi(x) \chi_{S}(x).
\end{equation}
Indeed, 
\begin{equation}
\langle \psi, \chi_S\rangle = \langle \sum_{T \subseteq [n]}\hat\psi(T)\chi_T, \chi_S\rangle = \sum_{T \subseteq [n]} \langle \hat\psi(T) \chi_T, \chi_S \rangle = \hat\psi(S),
\end{equation}
where the last equality follows from the orthonormality property.

For Fourier analysis involving Boolean functions, the typical convention is to consider function outputs in the set $\{-1, 1\}$ as opposed to $\{0, 1\}$.
When analyzing the Fourier coefficients of a binary function $\psi$ with the domain $\{0, 1\}$, we analyze the function $(-1)^{\psi(x)}$ rather than $\psi$ directly. For an in-depth discussion on Fourier analysis, refer to~\cite{odonnell}.
\section{Bounds on Cardinality of Set Sum}
\label{app:set-sum}
In this section, we prove Theorem~\ref{thm:sumset_bound}. The core idea of the proof is in the procedure that iteratively moves elements from the set $B$ to the set $A$ while shifting them by some vector. This operation doesn't change the sum of the sizes of the sets and is done in a manner that ensures the sumset cardinality at any given iteration is at most the cardinality of the sumset from the previous iteration. We provide a lower bound on the size of the sumsets when algorithm finishes and argue that initial subset's size is at least as large.

\begin{lemma}
    \label{lemma:sum_of_sets_strictly_larger_summand1}
    Let $A$ and $B$ be non-empty subsets of $\{0, 1\}^n$. If $B$ is not contained in any coset of a proper subspace of $\{0, 1\}^n$ and $A$ isn't equal to $\{0, 1\}^n$, then $|A + B| > |A|$.
    \begin{proof}
        In case $|B| > |A|$ the statement of the lemma is obvious because sets are nonempty. From now on we assume that $|A| \geq |B|$.
        
        For the sake of contradiction, assume that $|A + B| = |A|$. Now, let's define the set $B'$ as $B$ shifted by an element $b$ from $B$, i.e., $B' = B + b$. The zero element is contained in $B'$ and the size of the sumset $A + B'$ equals that of $A + B$, which in turn is $|A|$. Indeed, adding the element $b$ to each element in the sumset results in a bijection between $A + B'$ and $A + B$. Now, as $A \subseteq A + B'$ (since the zero element is in $A'$) and the sizes of the two sets are equal, we deduce that $A + B' = A$. Consequently, for every element $b' \in B'$, $b' + A = A$.
        
        Let's define a set $Q$ as the set of all elements $q$ in $\{0,1\}^n$ such that $q + A = A$. This set $Q$ satisfies the properties of a subspace of $\{0, 1\}^n$. Indeed, for any two elements $q_1$ and $q_2$ in $Q$, their sum when added to $A$ remains $A$, i.e., $q_1 + q_2 + A = q_1 + A = A$. However, $Q$ is not equal to $\{0, 1\}^n$. To illustrate this, for a given element $a_0$ in $A$, when $q$ varies over $\{0, 1\}^n$, the summation $q + a_0$ ranges over all elements in $\{0, 1\}^n$, which inevitably includes elements outside of $A$. Since every shifted set $b' + A$ with $b' \in B'$ is $A$, we have $B' \subseteq Q$. This implies that $B$ is contained in the coset defined by $b + Q$, leading to a contradiction, which finishes the proof of the lemma.
    \end{proof}
\end{lemma}

\begin{lemma}
    \label{lemma:sum_of_sets_strictly_larger_summand2}
    Let $A$ and $B$ be non-empty subsets of $\{0, 1\}^n$. Assume that $A$ is not contained in any coset of a proper subspace of $\{0, 1\}^n$. Let $Q$ be the smallest subspace of $\{0, 1\}^n$ such that $B$ is contained in a coset of $Q$. Then either $|A + B| > |A|$ or $A$ satisfies the following condition: for each coset of $Q$, either all vectors from that coset belong to $A$ or none do.
    \begin{proof}
        Let us consider the cosets of $Q$. For each coset, we select an arbitrary vector $q_i$ from that coset. Assume $B$ is contained in the coset $Q + \tilde{q}$. We define $A_i = A \cap (Q + q_i)$, that is, $A_i$ consists of the vectors from $A$ that are in the coset $Q + q_i$. We first prove that for distinct $A_i$ and $A_j$, their respective sum-sets $A_i + B$ and $A_j + B$ do not intersect. Consider arbitrary vectors $a_1 \in A_i, a_2 \in A_j$, and $b_1, b_2 \in B$. Notice that $a_1 + b_1 = q_i + \tilde{q} + (a_1 + q_i) + (b_1 + \tilde{q})$ and $a_2 + b_2 = q_j + \tilde{q} + (a_2 + q_j) + (b_2 + \tilde{q})$. As vectors $(a_1 + q_i), (a_2 + q_j), (b_1 + \tilde{q}), (b_2 + \tilde{q})$, belong to $Q$ and vectors $q_i$ and $q_j$ are from different cosets of $Q$, it follows that $a_1 + b_1$ and $a_2 + b_2$ must belong to different cosets, ensuring that $ (A_i + B) \cap (A_j + B) = \varnothing $.
        Consequently, the sum-set $A + B$ can be partitioned as follows:
        \begin{equation}
        A + B = \bigsqcup_{i} (A_i + B).
        \end{equation}

        We further note that $|A_i + B| = |A_i + q_i + B + \tilde{q}|$. Indeed XORing each element with $q_i + \tilde{q}$ establishes a bijection between these two sets. Since both $A_i + q_i$ and $B + \tilde{q}$ are contained in the subspace $Q$, and given that $Q$ is the smallest subspace containing a coset of $B$, Lemma~\ref{lemma:sum_of_sets_strictly_larger_summand1} can be applied unless $A_i + q_i$ is a empty or equal to $Q$. This results in $|A_i + B| > |A_i|$, unless $A_i$ is empty or contains all the vectors from corresponding coset. Combining this result with our partition of $A + B$ completes the proof.
    \end{proof}
\end{lemma}

Now we provide the main algorithm (see Algorithm~\ref{alg:the-only-one}).

\begin{algorithm}[t]
\textbf{Input:} $A_0, B_0$.
\begin{algorithmic}[1]
\State $i \gets 0$
\State $Q_0 \gets$ smallest subspace of $\{0, 1\}^n$ such that $B_0$ is contained in a coset of $Q_0$
\While{$\exists q \in \{0, 1\}^n: A_{i} \cap Q_{i} + q \neq 0, A_{i} \cap Q_{i} + q \neq Q_{i} + q$}
\State $b' \gets$ select arbitrary $b'$ in $B_i$
\State $\tilde{B} \gets B_i + b'$
\State $a' \gets$ select any $a'$ such that $a' + \tilde{B} \not\subseteq A_i$ \Comment{We can find such $a'$ by Lemma~\ref{lemma:sum_of_sets_strictly_larger_summand2}} 
\State $B' \gets \{b \in \tilde{B} | a' + b \notin A_i\}$
\State $A_{i + 1} \gets A_i \cup (a' + B')$
\State $B_{i + 1} \gets \tilde{B} \setminus B'$
\State $Q_{i + 1} \gets$ smallest subspace of $\{0, 1\}^n$ such that $B_{i + 1}$ is contained in a coset of $Q_{i + 1}$
\State $i \gets i + 1$
\EndWhile
\end{algorithmic}
\caption{Algorithm for Lemma~\ref{lemma:sumset_algorithm_invariants}}
\label{alg:the-only-one}
\end{algorithm}

\begin{lemma}
\label{lemma:sumset_algorithm_invariants}
Let $A_0, B_0, A_i, B_i$ be as given in Algorithm~\ref{alg:the-only-one}. The size of the setsum $A_0 + B_0$ is at least as large as that of $A_i + B_i$ at any iteration $i$ of the algorithm, and sizes of sets $A_0, B_0, A_i, B_i$ satisfy $|A_0| + |B_0| = |A_i| + |B_i|$ at each iteration.
\begin{proof}
We start by observing that $|A_i + B_i| = |A_i + \tilde{B}|$. This equality holds because $A_i + \tilde{B} = A_i + B_i + b'$, and XORing with $b'$ establishes a bijection between $A_i + B_i$ and $A_i + \tilde{B}$. The loop's condition assures us that there exists a coset of $Q_i$ such that its intersection with $A$ is neither empty nor consists of all vectors of the coset. Given that $\tilde{B}$ is simply $B$ translated by a vector $b$, $Q_i$ is also the smallest subspace of $\{0, 1\}^n$, coset of which contains $\tilde{B}$. Therefore, we can apply Lemma~\ref{lemma:sum_of_sets_strictly_larger_summand2} to conclude that $|A_i + \tilde{B}| > |A_i|$. This allows us to choose a vector $a'$ such that $a' + \tilde{B}$ is not a subset of $A_i$. By the definition of $a'$, $B'$ is non-empty. Now we construct the sets $A_{i+1}$ and $B_{i+1}$. They have the following properties:
First, $|A_{i+1}| = |A_i| + |B'|$. This is true because $A_i \cap (a' + B') = \varnothing$, which follows directly from the choice of $B'$. The cardinality of $\tilde{B} \setminus B'$ is $|B_i| - |B'|$. Consequently, $|A_{i+1}| + |B_{i+1}| = |A_i| + |B_i|$.

Next, $A_{i+1} + B_{i+1} \subseteq A_i + \tilde{B}$. The set $A_i + B_{i+1}$ is obviously contained in $A_i + \tilde{B}$. It remains to show that $(a' + B') + B_{i+1} = (a' + B') + (\tilde{B} \setminus B')$ is also contained in $A_i + \tilde{B}$. To demonstrate this, consider an arbitrary $a \in (a' + B')$ and $b \in \tilde B \setminus B'$. Then $a = a' + b'$ for some $b' \in B'$. Because $b$ is not in $B'$, $a' + b$ is an element of $A_i$. Therefore, $(a' + b) + b'$ belongs to $A_i + \tilde{B}$.
By induction, we conclude that $|A_i| + |B_i| = |A_0| + |B_0|$ and $|A_i + B_i| \leq |A_0 + B_0|$.
\end{proof}
\end{lemma}

It remains to prove the lower bound of the sumset size $|A_i + B_i|$ for the termination step of the algorithm. Initially, we construct $\tilde{B}$ to always include the element $0$ to ensure that $B_i$ is never empty throughout the algorithm. Indeed, if $B_i$ were empty at some iteration $i$, it would imply that $B' = \tilde{B}$ in the previous iteration $i-1$, which contradicts the fact that $a' + 0 \in A_i$ and therefore $0 \notin B'$. Consequently, $|A_i + B_i| \geq |A_i|$.

The algorithm halts when the condition specified in line 3 is not met. Specifically, given that $Q_i$ is the smallest subspace of $\{0, 1\}^n$ such that $B_i$ is contained in a coset of $Q_i$, for all cosets of $Q_i$, the intersection of $A_i$ with that coset is either empty or contains the entire coset. It follows that $|B_i| \leq |Q_i| = 2^{\dim Q_i}$, yielding
\begin{equation}
|A_0 + B_0| \geq |A_i + B_i| \geq |A_i| \geq |A_0| + |B_0| - |B_i| \geq |A_0| + |B_0| - 2^{\dim Q_i}.
\end{equation}

If the dimension of $Q_i$ is at most $2^{n-3}$, we obtain the desired bound. Next we consider the case when $\dim Q_i \geq 2^{n-2}$. We use the fact that $A_i \supseteq A_0$. When $\dim Q_i = n$ or $\dim Q_i = n-1$, it's straightforward to see that $A_i$ would span the entire $\{0, 1\}^n$ space. In the first case it follows since $A_0$ is non-empty and in the second case it follows because $A_0$ is not contained in neither $Q_i$, nor $\overline{Q_i}$.
Next, consider the case $\dim Q_i = n-2$. In this case, $Q_i$ has four distinct cosets. Since $A_0$ is not contained in any proper subspace of $\{0, 1\}^n$, it must contain elements in at least three of these cosets. Therefore, for these three cosets, $A_i$ would contain all the elements, leading to a size of $3 \times 2^{n-2}$ at the minimum. This concludes the proof of Theorem~\ref{thm:sumset_bound}.

\end{document}